\documentclass[%
 reprint,
 amsmath,amssymb,
aps,
pra,
]{revtex4-1}

\usepackage{graphicx}
\usepackage{dcolumn}
\usepackage{bm}

\newcommand{\formref}[1]{(\ref{#1})}
\newcommand{\brac}[1]{\left(#1\right)}

\newcommand{\bracr}[1]{\left[#1\right]}
\newcommand{\bracm}[1]{\left\langle #1\right\rangle}
\newcommand{\abs}[1]{\left|#1\right|}

\newcommand{\cet}[1]{\left|#1\right\rangle}

\begin{document}

\preprint{APS/123-QED}

\title
{Squeezing of thermal fluctuations in four-wave mixing in a \(\Lambda\)-scheme}


\author{Maria Erukhimova}
\email{eruhmary@appl.sci-nnov.ru}
\author{Mikhail Tokman}%
\affiliation{Institute of Applied Physics RAS, Uljanova str, 46,
Nizhny Novgorod, Russia}

\date{\today}

\begin{abstract}

We theoretically investigate the mechanism of two-mode quadrature squeezing in regime of four-wave mixing in a \(\Lambda\)-scheme of three-level atoms embedded in a thermal reservoir. We demonstrate that the process of nonlinear transfer of noise from the low frequency of ground state splitting to the optical frequency drastically modifies the condition of effective two-mode squeezing. The damage factor is significant if number of thermal photons at the low frequency is high and the role of inelastic processes in ground state coherence decay is not negligible. We found the optimal conditions for squeezing, in particular optimal density-length product of active medium depending on the relaxation parameters and drive intensity.

\begin{description}
\item[PACS numbers] 42.50.Gy, 42.50.Lc, 42.65.-k

\end{description}
\end{abstract}

\pacs{Valid PACS appear here}
\maketitle

\section{\label{Introduction}Introduction}

Since the pioneering work of Slusher \cite {Slusher} many groups have demonstrated squeezing based on four-wave mixing (4WM) in atomic vapors under a variety of different conditions \cite{Maeda, Vallet, Lambrecht, Josse}. A number of theoretical and experimental works \cite{Grove, Shahriar, Lukin_Matsko, Balic, Kolchin06, Kolchin07, McCormickOL, McCormick, BoyerScience, Glorieux} have shown that resonant 4WM based on ground-state coherence is attractive, since coherent population trapping (CPT) and electromagnetically induced transparency (EIT) can provide small ratio of absorption coefficient with respect to high increment of 4WM instability, and as a consequence strong intensity correlations of generated twin beams and high level of two-mode quadrature squeezing between opposite sidebands of the twin beams. 
One of advantages of this mechanism is that 4WM naturally generates the squeezed light with narrow frequency band that is resonant to an atomic transition.  So that it can be applied as a quantum-information carrier interacting with a material system. And, besides, in resonant schemes of 4WM high level of squeezing can be achieved for relatively low drive powers.   
Noise correlations and squeezed light generation on the basis of 4WM in sodium vapor was experimentally demonstrated in papers \cite{Grove}. The demonstration of \(-8 dB\) of relative intensity squeezing between probe and conjugate beams produced in nearly copropagating nondegenerate 4WM scheme in rubidium vapor  was reported in \cite{McCormickOL, McCormick, Qin}.

A common problem in experiments using resonant atomic ensembles for squeezing is the processes of spontaneous emission in atoms that occur concurrently to four-wave mixing and lead to incoherent emission into the signal modes. These processes contribute to the ``thermalization'' of the optical state in the signal modes and degrade the squeezing \cite{Lvovsky}. Such processes take place if there are thermal excitations in the medium (if the temperature is not zero). 

The 4WM squeezing experiments and experiments on paired photon generation in both ``cold'' \cite{Balic, Kolchin06} and ``hot'' \cite{McCormick} atomic vapours were performed.   
Most of theoretical treatments \cite{Lukin_Matsko, Glorieux} use so called ground-state approximation (or zero-temperature approximation) for analysis of squeezed light generation in such schemes. This approximation appeals to the CPT effect, so that under condition of negligible relaxation at the ground state transition the most of atoms are ``trapped'' in the ground state even if the thermal excitation energy far exceeds the ground state splitting frequency. The generally excepted idea \cite{McCormickOL} is that the negative influence of spontaneous emission on squeezing in 4WM based on ground-state coherence is reduced due to the EIT effect. It is intuitively based on the standard fluctuation-dissipation theorem (FDT) \cite{Callen}.  

But the generalized fluctuation-dissipation relation obtained in \cite{ErukhOL15} for a resonantly driven quantum medium makes us to believe that the situation is complicated by the fact that the noise level at the optical frequency may be determined by the averaged number of thermal photons at the low frequency of ground state splitting if the rate of spontaneous emission at this frequency is not zero.  This effect is caused by the parametric transfer of noise from low to high frequency in the presence of resonant drive wave. Actually the generalized FDT \cite{ErukhOL15} takes into account the process of spontaneous Raman scattering that leads to incoherent emission into the resonant signal mode and becomes notable under condition of unideal CPT. 

The presence of some extra noise beyond that which is necessary to preserve the canonical commutation relation of the field was observed in the EIT delayed light experiments \cite{Hsu} and in experiments on the propagation of squeezed vacuum under EIT \cite{Figueroa}, and was also associated with the processes of population exchange between ground states in theoretical treatment \cite{Hetet}. The calculation of spontaneous Raman noise was presented in paper \cite{Reim} demonstrated storage and retrieval of single photons in off-resonant Raman memory scheme.  

The importance of noise scattering in degradation of squeezing in 4WM mechanisms was pointed in \cite{Lvovsky}. But until now the accurate theory of 4WM squeezed light generation in ``hot'' atomic vapours  based not on the phenomenological approach (as in \cite{McCormick}), but with gain, propagation losses and noise intrinsically included in the microscopic approach has not been presented.

In the present paper we theoretically investigate the process of generation of two-mode squeezed vacuum based on 4WM mechanism realized in three-level \(\Lambda\)-scheme  with monochromatic driving under condition of thermal excitations. The considered system is interesting also from the methodological point of view as a spectacular example that illustrates the mechanisms 
of transformation of the field noise due to the self-consistent parametric interaction in resonant nonlinear medium.  

We use the results of previous research of this system \cite{Vdovin} conducted for the case of zero temperature and negligible dissipation, as well as results of analysis of this scheme of parametric instability for classical fields \cite{Tokman, ErukhPRA11}. 

In section \ref{Formalism} we introduce with the necessary degree of detail the basic terms and equations used for analysis of quantum radiation propagating through the dense medium consisting of independent quantum centers (``atoms'') interacted with dissipative reservoir under condition of nonzero temperature.
In section \ref{scheme} the system of resonant 4WM in a \(\Lambda\)-scheme is described. The characteristics of parametric interaction of waves taking into account the thermal redistribution of populations among energy levels are obtained. On the basis of approach developed in paper \cite{ErukhOL15} the correlation relations for the noise sources of interacted quantum fields are derived. 
In section \ref{squeezing} the regime of two-mode squeezed vacuum generation in this system is analysed. The spectral properties of the squeezed vacuum at different temperatures are investigated. The optimal parameters for squeezing are obtained.


\section{\label{Formalism} General equations}

Here we introduce in detail the basic terms and relations for the system of quantum radiation interacting with the medium consisting of quantum atoms embedded in a dissipative reservoir in frame of Heisenberg-Langevin approach. 
  
\subsection{Atomic system}
For the atomic system we define the coordinate-dependent density operator as
\begin{equation}
\hat{\rho}_{mn}\brac{\mathbf{r},t}=\frac{1}{\Delta V_r}\sum_j\hat{\rho}_{j;mn}(t),
\end{equation}
where index \(j\) numerates the atoms within the small volume \(\Delta V_r\) in the vicinity of point with radius-vector \(\mathbf{r}\), \(\hat{\rho}_{j;mn}=\hat{a}_{j;n}^{\dag}\hat{a}_{j;m}\) is the Heisenberg density operator acting on variables of atom with index \(j\); it is expressed via the creation and annihilation operators 
which are defined by expressions \(\hat{a}_{n}^{\dag}\cet{0}=\cet{n}\), \(\hat{a}_{n}\cet{n}=\cet{0}\), where \(\cet{n}\) are basis states of single-particle Hamiltonian with energy levels \(W_n\). 
The operator \(\hat{\rho}_{mn}\) obeys the Heisenberg-Langevin equation \cite{Annals, Tokman_Yao_Belyanin}: 
\begin{equation}
\dot{\hat{\rho}}_{mn}
=-\frac{i}{\hslash}\brac{\hat{h}_{mp}\hat{\rho}_{pn}-\hat{\rho}_{mp}\hat{h}_{pn}}+\hat{R}_{mn}+\hat{F}_{mn},
\label{rho}
\end{equation}  
where \(\hat{h}_{mn}=W_m\delta_{mn}-\mathbf{d}_{mn}\hat{\mathbf{E}}\brac{\mathbf{r},t}\) takes into account interaction of atoms with electric field \(\hat{\mathbf{E}}\brac{\mathbf{r},t}\) in the dipole approximation, \(\mathbf{d}_{mn}\) is the dipole matrix elements, \(\hat{R}_{mn}\) is the relaxation operator, and \(\hat{F}_{mn}\) is the Langevin noise operator satisfying \(\hat{F}_{mn}=\hat{F}^\dag_{nm}\), \(\bracm{\hat{F}_{mn}}=0\), hereinafter averaging is taken over the reservoir variables and atomic state.

The standard model for the relaxation operators  corresponds to the so-called master equations \cite{Blum, Fain, Scully}, where 
\begin{equation}
\hat{R}_{mn}=\sum_{pq}r_{mnpq}\hat{\rho}_{pq}.
\label{r_mnpq}
\end{equation} 
In a simplest form the nonzero coefficients \(r_{mnpq}\) set the rates of transverse and longitudinal relaxation:
\begin{eqnarray}
\hat{R}_{mn}=-\gamma_{mn}\hat{\rho}_{mn}, m\neq n, \nonumber \\
\hat{R}_{mm}=\sum_n{w_{mn}\hat{\rho}_{nn}}.
\label{gamma_mn}
\end{eqnarray} 
The correlation functions for the atomic noise operators are derived in Appendix~\ref{Appendix2}.  Here we'll use the correlation relations for the spectral components of the Langevin operators, defined as \(\hat{F}_{mn}\brac{\mathbf{r},t}=\int_{-\infty}^{+\infty}{\hat{F}_{mn}(\mathbf{r},\omega)e^{-i\omega t}d\omega}\). Under the adiabatic approximation neglecting the slow evolution of populations \(\bracm{\hat{\rho}_{nn}}\) and amplitude of drive-induced coherence at resonant transition \(\sigma_{ba}\), defined as \(\bracm{\hat{\rho}_{ba}}=\left.{\sigma_{ba}e^{\mp i\omega_d t}}\right|_{b\gtrless a}\), we get:
\begin{eqnarray}
\bracm{\hat{F}_{mn}(\mathbf{r},\omega)\hat{F}_{nm}(\mathbf{r'},\omega')}=\nonumber\\
\frac{1}{2\pi}\brac{2\gamma_{mn}\bracm{\hat{\rho}_{nn}}+\bracm{\hat{R}_{nn}}}\delta(\omega+\omega')\delta(\mathbf{r}-\mathbf{r}')\nonumber\\
\label{mnnm_omega}
\end{eqnarray}
\begin{eqnarray}
\bracm{\hat{F}_{ma}(\mathbf{r},\omega)\hat{F}_{bm}(\mathbf{r'},\omega')}=
\frac{1}{2\pi}\brac{\gamma_{am}+\gamma_{bm}-\gamma_{ab}}\times\nonumber\\
\left.{\sigma_{ba}\delta(\omega+\omega'\mp \omega_d)}\right|_{b\gtrless a}\delta(\mathbf{r}-\mathbf{r}').\nonumber\\
\label{mabm_omega}
\end{eqnarray}

\subsection{The field equations\label{The field equations}}

The electric field operator \(\hat{\mathbf{E}}\brac{\mathbf{r},t}\) in a general case obeys the following operator wave equation \cite{Annals, Fain}:
\begin{equation}
\frac{\partial^2}{\partial t^2}\brac{\hat{\mathbf{E}}+4\pi \hat{\mathbf{P}}}+c^2\nabla\times\nabla\times\hat{\mathbf{E}}=0,
\label{fulleq}
\end{equation}
where the operator of electric polarization \(\hat{\mathbf{P}}\brac{\mathbf{r},t}=\sum_{m,n}\mathbf{d}_{nm}\hat{\rho}_{mn}\) is expressed via density operators \(\hat{\rho}_{mn}\), which are the solution of the Eq.~\formref{rho}. Assume field to be quantized over spatial modes in homogeneous, dissipationless medium with linear dielectric permittivity \(\epsilon(\omega)=1+\int_0^\infty4\pi\chi^H(\tau)e^{i\omega \tau}d\tau\), whereas the nonlinear response, dissipation and noise effects are taken into account as small additional term to the linear relation:
\begin{equation}
\hat{\mathbf{P}}\brac{\mathbf{r},t}=\int_0^{\infty}\chi^H(\tau)\hat{\mathbf{E}}\brac{\mathbf{r},t-\tau}d\tau+\delta\hat{\mathbf{P}}.
\label{P}
\end{equation}
Here \(\chi^{H}\) is the Hermitian part of the linear susceptibility of the medium.
We assume that the field consists of one or several quasi-monochromatic waves labelled by index \(j\), each of them may be presented as combination of large number of spectrally close modes of quantization propagating within a paraxial beam of a cross-sectional area \(S_\perp\). Then the following representation can be used 
\begin{eqnarray}
\hat{\mathbf{E}}\brac{\mathbf{r},t}=\nonumber\\
\sum_{j}{\mathbf{e_j}E_j\hat{c}_j\brac{\mathbf{r},t}e^{i\mathbf{k_j}\mathbf{r}-i\omega_j t}
+\mathbf{e_j}^*E_j\hat{c}_j^\dag\brac{\mathbf{r},t}e^{-i\mathbf{k_j}\mathbf{r}+i\omega_j t}},\nonumber
\end{eqnarray}
where \(\mathbf{e_j}\) is the unit vector of the polarisation of \(j\)-wave, \(\omega_j\) and \(\mathbf{k_j}\) are coupled by corresponding dispersion relation \(k_j^2c^2/\omega_j^2=\epsilon(\omega_j)\), operators \(\hat{c}_j\brac{\mathbf{r},t}\) and \(\hat{c}_j^\dag\brac{\mathbf{r},t}\) are slowly time- and space- dependent photon annihilation and creation operators. \(E_j\) are normalization constants defined as \({E_j}=\sqrt{4\pi\hbar\omega_j^2/\abs{\frac{\partial\brac{\omega_j^2\epsilon(\omega_j)}}{\partial\omega_j}}}\)  \cite{Fain, Annals}. For such representation the operators \(\hat{n}_{phj}=\hat{c}_j^\dag\brac{\mathbf{r},t}\hat{c}_j\brac{\mathbf{r},t}\) 
play the role of the photon density operators and \(\mathbf{\hat{p}}_{phj}=\mathbf{v}_{grj}\hat{c}_j^\dag\brac{\mathbf{r},t}\hat{c}_j\brac{\mathbf{r},t}\) are the photon flux density operators, where \(\mathbf{v}_{grj}=2c^2\mathbf{k_j}/\frac{\partial\brac{\omega_j^2\epsilon}}{\partial\omega_j}\) are the group velocities.

The following truncated equations can be written for the operators \(\hat{c}_j\brac{\mathbf{r},t}\) \cite{Vdovin, Tokman_Yao_Belyanin}: 
\begin{equation}
\brac{\frac{\partial}{\partial t}+\brac{\mathbf{v}_{grj}\nabla}}\hat{c}_j\brac{\mathbf{r},t}=\Lambda_j\frac{i}{\hbar}\mathbf{e}^*E_j\delta\hat{\mathbf{P}}_{j}\brac{\mathbf{r},t},
\label{c_eq}
\end{equation}  
where \(\Lambda_j=sign\brac{\frac{\partial\brac{\omega_j^2\epsilon}}{\partial\omega_j}}\), \(\delta\hat{\mathbf{P}}_{j}\brac{\mathbf{r},t}\) are the slowly varying amplitudes of the polarization terms \(\delta\hat{\mathbf{P}}=\sum_j{\delta\hat{\mathbf{P}}_{j}\brac{\mathbf{r},t}e^{i\mathbf{k_j}\mathbf{r}-i\omega_j t}
+\delta\hat{\mathbf{P}}_{j}^\dag\brac{\mathbf{r},t}e^{-i\mathbf{k_j}\mathbf{r}+i\omega_j t}}\).

In boundary-value problem the spectral decomposition of the field amplitude is used:
\begin{equation}
\hat{c}_j\brac{\mathbf{r},t}=\int_{-\infty}^{+\infty}\hat{c}_{j_\nu}\brac{\mathbf{r}}e^{-i\nu t}d\nu ,
\label{spectral}
\end{equation} 
where the integration in infinite limits is reduced to the integration over narrow frequency interval \(\Delta \nu_j\ll\omega_j\). It is appropriate to write the propagation equation for the ``flux" amplitudes \(\hat{s}_j=\sqrt{\abs{v_{grj}}}\hat{c}_j\). Assuming that the wave propagates in positive \(z\) direction,  \(\mathbf{k_j}=k_j\mathbf{z_0}\), we get the following equation:
\begin{equation}
\frac{\partial \hat{s}_{j_\nu}\brac{z}}{\partial z}-\frac{i\nu}{v_{grj}}\hat{s}_{j_\nu}\brac{z}=\frac{i\mathbf{e}^*E_j}{\hbar\sqrt{\abs{v_{grj}}}}\delta\hat{\mathbf{P}}_{j_\nu}\brac{z},
\label{prop}
\end{equation}  
where \(v_{grj}=2c^2k_j/\frac{\partial\brac{\omega_j^2\epsilon}}{\partial\omega_j}\), \(\delta\hat{\mathbf{P}}_{j_\nu}\brac{\mathbf{r}}\) is the corresponding spectral image of the slowly varying amplitude of polarization:
\begin{equation}
\delta\hat{\mathbf{P}}_j\brac{\mathbf{r},t}=\int_{-\infty}^{+\infty}\delta\hat{\mathbf{P}}_{j_\nu}\brac{\mathbf{r}}e^{-i\nu t}d\nu .
\end{equation}
At the boundary \(z_b\) between the medium and vacuum the following boundary condition for the field operator can be used: 
\begin{equation}
\frac{1}{\sqrt{c}}\hat{s}_j(z_b)\mid_{medium}=\hat{c}_j(z_b)\mid_{vacuum},
\label{boundary}
\end{equation}
which satisfies the conservation of Poynting flux. (It is assumed that the effects of reflection are neglected.) 

The polarization on the right-hand side of Eq.~\formref{prop} includes nonlinearity (e.g. parametric coupling of different waves), dissipation and fluctuations: \(\delta\hat{\mathbf{P}}_{j_\nu}=\delta\hat{\mathbf{P}}_{j_\nu}^{N}+\delta\hat{\mathbf{P}}_{j_\nu}^{diss}+\delta\hat{\mathbf{P}}_{j_\nu}^{L}\). The dissipation term that is linear over quantum field: \(\delta\hat{\mathbf{P}}_{j_\nu}^{diss}=\chi^{aH}(\omega_j+\nu)\mathbf{e}E_j\hat{c}_{j_\nu}\), defines the absorption coefficient \(\kappa_j(\nu)=-i\frac{2\pi k_j}{\epsilon_j}\chi^{aH}(\omega_j+\nu)\). Here \(\chi^{aH}\) is the anti-Hermitian part of the susceptibility. Note, that if the medium is driven by classical field the ``linear'' properties depends on drive intensity as on the parameter \(\epsilon(I_d)\), \(\chi^{aH}(I_d)\), \(\kappa_j(I_d)\). 

So we use the following form of equation for the filed operator:
\begin{equation}
\frac{\partial \hat{s}_{j_\nu}\brac{z}}{\partial z}-\frac{i\nu}{v_{grj}}\hat{s}_{j_\nu}\brac{z}+\kappa_j(\nu)\hat{s}_{j_\nu}\brac{z}=\hat{N}_{j_\nu}+\hat{L}_{j_\nu}.
\label{prop2}
\end{equation}  
Here \(\hat{N}_{j_\nu}=\frac{i\mathbf{e}^*E_j}{\hbar\sqrt{\abs{v_{grj}}}}\delta\hat{\mathbf{P}}_{j_\nu}^N\), \(\hat{L}_{j_\nu}=\frac{i\mathbf{e}^*E_j}{\hbar\sqrt{\abs{v_{grj}}}}\delta\hat{\mathbf{P}}_{j_\nu}^L\). 

The Langevin term \(\hat{L}_{j_\nu}\) provides the fulfilment of commutation relations for the field operators in a dissipative medium \cite{Vdovin, Tokman_Yao_Belyanin}. 
\begin{eqnarray}
\bracr{\hat{s}_{j_\nu}(z),\hat{s}_{j_{\nu'}}^\dag(z)}=\frac{1}{2\pi S_\bot}\delta\brac{\nu-\nu'}.
\nonumber
\end{eqnarray}
So the definite relation between noise terms and absorption coefficient follows from this general requirement without specification of the dissipation and noise origin:  
\begin{equation}
\bracr{\hat{L}_{j_\nu}\brac{z},\hat{L}_{j_{\nu'}}^\dag\brac{z'}}=\frac{1}{\pi S_\bot}\kappa_{j}(\nu)\delta\brac{z-z'}\delta\brac{\nu-\nu'}.
\label{commutatorLnu3} 
\end{equation} 
But the separate expressions for the correlation functions of Langevine operators \(\bracm{\hat{L}_{j_\nu}\brac{z}\hat{L}_{j_{\nu'}}^\dag\brac{z'}}\), \(\bracm{\hat{L}_{j_{\nu'}}^\dag\brac{z'}\hat{L}_{j_\nu}\brac{z}}\) can not be evaluated from this relation. Their correct calculation should be based on the ``microscopic'' approach, by consistent calculation of noise atomic response on the action of atomic noise operators \(\hat{F}_{mn}\), obeying correlation relations \formref{mnnm_omega}, \formref{mabm_omega}, and corresponding evaluation of noise component of medium polarization.

\section{\label{scheme} Resonant four-wave mixing in a \(\Lambda\)-scheme}
\subsection{The key parameters of the model \label{key parameters}}
\begin{figure}
\includegraphics[width=\columnwidth, keepaspectratio=true]{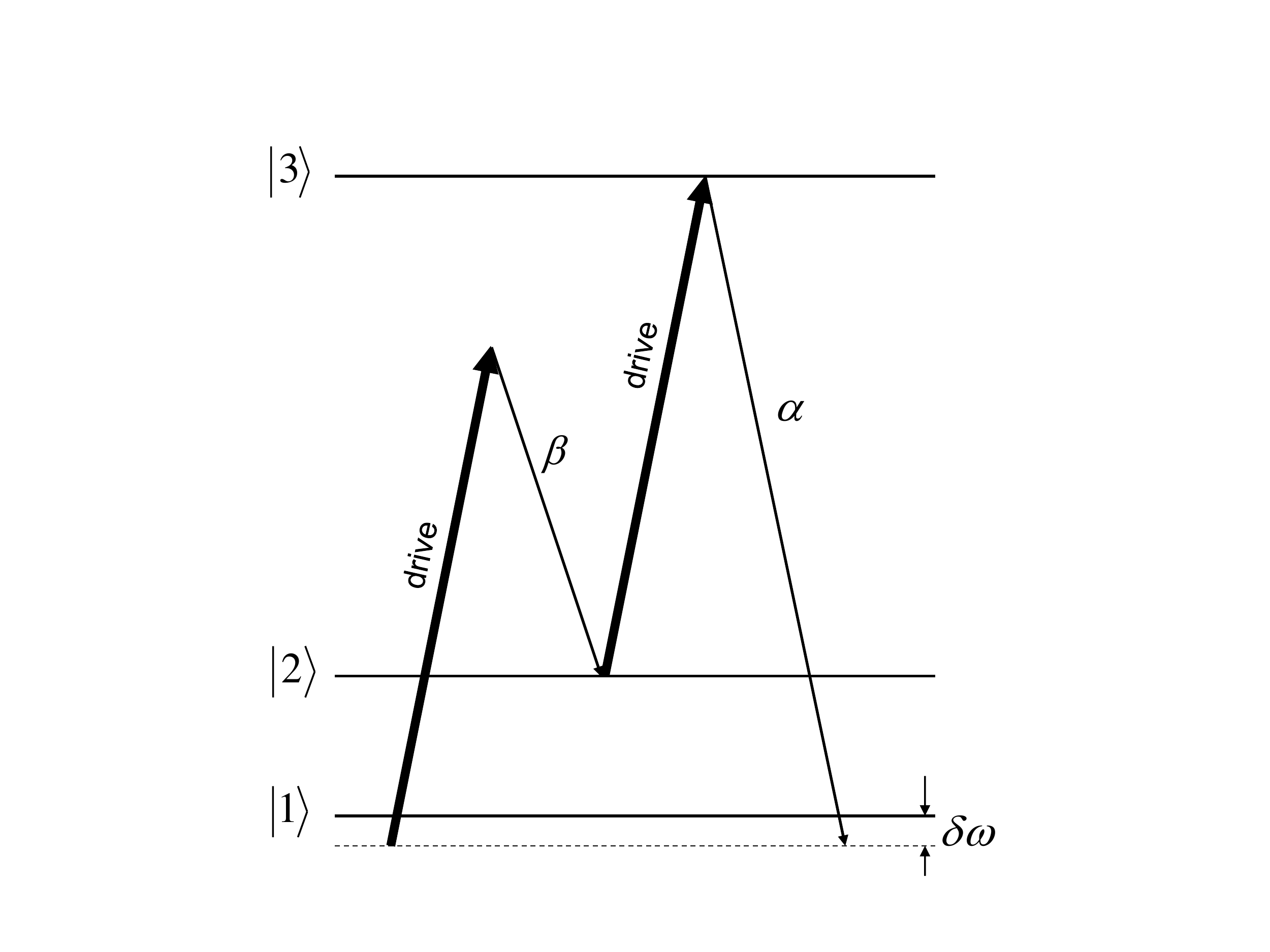}
\caption{\label{figure1} The scheme of resonant four-waves mixing process in three level \(\Lambda\)-atoms.}
\end{figure} 
We consider the following scheme of four-wave mixing process, depicted in Fig.~\ref{figure1}, where the total field consists of the strong classical drive wave and signal quantum waves: \(\alpha\)-wave and \(\beta\)-wave, co-propagating in a \(z\)-direction throw the plane-parallel layer of three-level atoms:
\begin{eqnarray}
\hat{\mathbf{E}}=\mathbf{e_d}\xi_d e^{ik_dz-i\omega_d t}+c.c.+\nonumber\\
\mathbf{e_\alpha}E_\alpha\hat{c}_\alpha\brac{z,t}e^{ik_\alpha z-i\omega_\alpha  t}+\mathbf{e_\beta}E_\beta\hat{c}_\beta\brac{z,t}e^{ik_\beta z-i\omega_\beta t}+H.c.\nonumber\\
\label{E}
\end{eqnarray}
For simplicity we put \(\mathbf{e_d}=\mathbf{e_\alpha}=\mathbf{e_\beta}=\mathbf{e}\).
The frequency of classical wave with constant amplitude \(\xi_d=\abs{\xi_d}^{i\theta}\) is close (for simplicity equal) to the the frequency of the atomic transition \(\cet{2}-\cet{3}\): \(\omega_d=\omega_{32}\). The spectral decomposition of the wave amplitudes is used Eq.~\formref{spectral}: 
\begin{equation}
\hat{c}_{\alpha,\beta}\brac{\mathbf{r},t}=\int_{-\infty}^{+\infty}\hat{c}_{\alpha,\beta_\nu}\brac{\mathbf{r}}e^{-i\nu t}d\nu.
\label{spectral1}
\end{equation} 

Summarizing the results of previous investigation of this system  \cite{Vdovin, Tokman} we have, that
the effective generation of the bichromatic correlated radiation (and squeezed two-mode radiation) at the frequencies obeying four-wave resonance condition: 
\begin{equation}
\omega_\alpha+\omega_\beta=2\omega_d
\end{equation}
is realized due to the combination of the following factors:

(i)  Electromagnetically Induced Transparency. 
The drive wave should be strong enough but can be far from saturation value:
\begin{eqnarray}
\gamma_{31}^2\gg\abs{\Omega}^2\gg\gamma_{31}\gamma_{21}, 
\label{EITcond}
\end{eqnarray}
where \(\Omega=\brac{\mathbf{d_{32}e_d}\xi_d}/2\hbar\) is the Rabi frequency of drive wave, \(\gamma_{ij}\)  are the coherence decay rates at corresponding atomic transitions.
The frequency of \(\alpha\)-wave should be close to the two-photon resonance (\(\omega_\alpha-\omega_d\approx\omega_{21}\)), that in our case corresponds to the resonance with the atomic transition \(\cet{1}-\cet{3}\), adjacent to the  transition \(\cet{1}-\cet{2}\) with long lived coherence \(\gamma_{21}\ll\gamma_{31},\gamma_{32}\). 
The width of EIT window is defined by condition \(\abs{\omega_\alpha-\omega_{31}}<\Delta_{EIT}\), where \(\Delta_{EIT}\sim\abs{\Omega}^2/\gamma_{31}\).
The EIT condition Eq.~\formref{EITcond} provides both sufficient reduction of the partial decrement of resonant \(\alpha\)-wave and strong parametric coupling for configuration with close lower atomic levels:  
\begin{equation}
\omega_{21}\ll\omega_{31},\omega_{32}.
\label{small21}
\end{equation}

(ii) Four-wave spacial synchronism. This condition is provided by favourable wave dispersion in the considered atomic medium. We suppose that the ``central'' frequencies of \(\alpha\) and \(\beta\) waves correspond to the strict four-wave spacial synchronism: 
\begin{equation}
k_\alpha+k_\beta=2k_d,
\end{equation} 
where \(k_j=(\omega_j/c)\sqrt{\epsilon(\omega_j)}\).
It was shown in \cite{Tokman, ErukhPRA11} that it is fulfilled for definite detuning from the resonance with atomic transition \(\omega_\alpha=\omega_{31}+\delta\omega\), where:
\begin{eqnarray}
\delta\omega=\frac{\abs{\Omega}^2}{\omega_{21}}\brac{\frac{3}{2}+\abs{\zeta}^2},\label{detune1}
\end{eqnarray}
here \(\zeta=\frac{\mathbf{d_{31}e}}{\mathbf{d_{32}e}}\). 

The effective generation of correlated radiation is realized if parametric coupling coefficient is large compared with the  partial decrement of \(\alpha\)-wave, that is reduced to the following relation:
\begin{eqnarray}
\frac{\abs{\Omega}^2}{\gamma_{21}\omega_{21}}>1.
\label{strong}
\end{eqnarray}
The smallness of the synchronism shift compared to the parametric coupling coefficient determines the frequency band of parametric instability near the central frequency \(\abs{\nu}<\Delta_P\), where
\begin{eqnarray}
\Delta_P=\frac{2\abs{\Omega}^2}{\omega_{21}}
\label{Delta_P}.
\end{eqnarray}

The regime of two-mode squeezed vacuum generation in this system was analysed in detail in paper \cite{Vdovin} for the case of cold atoms and negligible dissipation. Our aim is to analyse this regime in a hot atomic medium, so we'll use the results of paper \cite{Vdovin} where they are applicable only.

\subsection{Equations for the density-matrix operators}
 
Here we write the equations for density-matrix operators Eq.~\formref{rho} in the presence of electric field  \formref{E} and Langevin forces \(\hat{F}_{mn}\) in order to calculate the coherently driven partial, parametric and noise components of polarisation. Using rotating wave approximation (RWA) with respect to high frequencies \(\omega_{31},\omega_{32}\) under condition Eq.~\formref{small21} the interaction of all field components with both high-frequency transitions is taken into account, so that the density-matrix operators are presented as:
\begin{eqnarray}
\hat{\rho}_{31}=\sum_{j=d,\alpha, \beta}\hat{\sigma}_{31}^je^{-i\omega_j t+ik_j z}, \hat{\rho}_{32}=\sum_{j=d,\alpha, \beta}\hat{\sigma}_{32}^je^{-i\omega_j t+ik_j z},\nonumber\\
\hat{\rho}_{21}=\hat{\sigma}_{21}e^{-i\omega_l t+ik_l z},\nonumber
\end{eqnarray}
where \(\omega_{l}=\omega_\alpha-\omega_d\), \(k_l=k_\alpha-k_d\). 
Meanwhile, we use the approximation of well resolved transitions  
\begin{eqnarray}
\gamma_{31},\gamma_{32}<<\omega_{21}.
\label{resolved}
\end{eqnarray} 

We also assume the following condition to be fulfilled:
\begin{eqnarray}
\frac{\gamma_{31}\abs{\Omega}^2}{\gamma_{21}\omega_{21}^2}\ll 1.
\label{small}
\end{eqnarray}
As it will be seen later, the condition Eq.~\formref{small} enables to neglect the partial dissipation of the nonresonant \(\beta\)-wave with respect to the dissipation of the resonant \(\alpha\)-wave. It is fulfilled if small relaxation rate at transition \(\cet{1}-\cet{2}\) is taken into account.

Consider the atomic response to the action of quantum radiation and noise forces additively in linear approximation over quantum fields. Assume that the amplitude of classical drive wave is constant.  Working in spectral representation for the field operators Eq.~\formref{spectral} we use corresponding representation for the slow components of the density-matrix operators:
\begin{eqnarray}
\hat{\sigma}_{31,32}^{\alpha,\beta}\brac{z,t}=\int_{-\infty}^{+\infty}\hat{\sigma}_{31,32_\nu}^{\alpha,\beta}\brac{z}e^{-i\nu t}d\nu,\nonumber \\
\hat{\sigma}_{21}\brac{z,t}=\int_{-\infty}^{+\infty}\hat{\sigma}_{21_\nu}\brac{z}e^{-i\nu t}d\nu,\nonumber
\end{eqnarray}
where the width of corresponding spectral lines are small compared to the atomic frequency \(\omega_{21}\). Similarly the spectrum of Langevin forces \(F_{mn}\brac{z,\omega}\) can be separated into three intervals
\begin{eqnarray}
\hat{F}_{31,32}(z,\omega)=\sum_{j=\alpha, \beta}\hat{f}_{31,32_{\nu=\omega-\omega_j}}^j\brac{z},\nonumber\\
\hat{F}_{21}\brac{\omega,z}=\hat{f}_{21_{\nu=\omega-\omega_l}}\brac{z}
\label{F_f}
\end{eqnarray}
Then the system of equations for the density-matrix operators takes the following form of algebraic equations:
\begin{eqnarray}
\hat{\sigma}_{31_\nu}^{\alpha}=\frac{1}{\Delta_{31}^\alpha}\brac{\hat{\alpha}_\nu\brac{\rho_{11}-\rho_{33}}+\Omega\hat{\sigma}_{21_\nu}+\hat{f}_{31_\nu}^\alpha}\nonumber\\
\hat{\sigma}_{31_\nu}^{\beta}=\frac{1}{\Delta_{31}^\beta}\brac{\zeta\hat{\beta}_\nu\brac{\rho_{11}-\rho_{33}}+\hat{f}_{31_\nu}^\beta}\nonumber\\
\hat{\sigma}_{32_\nu}^{\alpha}=\frac{1}{\Delta_{32}^\alpha}\brac{\frac{1}{\zeta}\hat{\alpha}_\nu\brac{\rho_{22}-\rho_{33}}+\hat{f}_{32_\nu}^\alpha}\nonumber\\
\hat{\sigma}_{32_\nu}^{\beta}=\frac{1}{\Delta_{32}^\beta}\brac{\hat{\beta}_\nu\brac{\rho_{22}-\rho_{33}}+\zeta\Omega\brac{\hat{\sigma}_{21_{-\nu}}}^\dag+\hat{f}_{32_\nu}^\beta}\nonumber\\
\hat{\sigma}_{21_\nu}=
\frac{1}{\Delta_{21}^l}\times\nonumber\\
\times\brac{\hat{\sigma}_{31_\nu}^\alpha\Omega^*+\sigma_{31}^d\brac{\hat{\beta}_{-\nu}}^\dag-\hat{\alpha}_\nu\sigma_{32}^{d*}
-\zeta\Omega\brac{\hat{\sigma}_{32_{-\nu}}^\beta}^\dag+\hat{f}_{21_\nu}}
\nonumber\\\label{sigma1}
\end{eqnarray}
Here \(\zeta=\frac{\mathbf{d_{31}e}}{\mathbf{d_{32}e}}\), as was introduced in Sec.~\ref{key parameters}. The field operators are
\begin{eqnarray}
\hat{\alpha}_\nu=\frac{\mathbf{d_{31}e}E_\alpha\hat{c}_{\alpha_\nu}}{2\hbar}, \hat{\beta}_{\nu}=\frac{\mathbf{d_{32}e}E_\beta\hat{c}_{\beta_\nu}}{2\hbar}. \nonumber
\end{eqnarray} 
The complex frequency detunings are:
\begin{eqnarray}
\Delta_{mn}^j=\omega_{mn}-\omega_j-\nu-i\gamma_{mn}, j=d,l,\alpha, \beta \label{Delta}
\end{eqnarray}
For the averaged drive-induced coherences we have the following relations:
\begin{eqnarray}
\sigma_{31}^{d}=\frac{1}{\Delta_{31}^d}{\zeta\Omega\brac{\rho_{11}-\rho_{33}}},\nonumber\\
\sigma_{32}^{d}=\frac{1}{\Delta_{32}^d}{\Omega\brac{\rho_{22}-\rho_{33}}},\label{sigma2}
\end{eqnarray}
where \(\Delta_{31}^d\approx\omega_{21}\), \(\Delta_{32}^d=-i\gamma_{32}\).
The averaged values for diagonal density-matrix operators obey the equations:
\begin{eqnarray}
-\brac{w_{21}+w_{31}}\rho_{11}+w_{12}\rho_{22}+w_{13}\rho_{33}=0\nonumber\\
-2Im\brac{\Omega^*\sigma_{32}^d}+w_{21}\rho_{11}-\brac{w_{12}+w_{32}}\rho_{22}+w_{23}\rho_{33}=0\nonumber\\
\label{rho1}
\end{eqnarray}
In Eqs~\formref{sigma1},\formref{Delta},\formref{sigma2},\formref{rho1} the relaxation model Eq.~\formref{gamma_mn} is used.

\subsection{Field-induced populations redistribution}

Under condition of finite temperature \(T\neq 0\) and not ideal low-frequency transition \(\cet{2}-\cet{1}\) it is necessary to calculate the redistribution of atoms over levels induced by the drive wave. To this end we express the relaxation rates in term of equilibrium (in the absence of drive) population distribution \(r_n^T=\rho_{nn}^T/\rho_{11}^T=exp\brac{-\hbar\omega_{n1}/T}\) and longitudinal relaxation times, defined for transition \(\cet{m}-\cet{n}\) as \(T_{mn}=\brac{A_{mn}\brac{n_T(\omega_{mn})+1}}^{-1}\), where \(A_{mn}\) are the Einstein coefficients, \(n_T(\omega)=\brac{e^{\hbar\omega/T}-1}^{-1}\) is the averaged number of ``thermal'' photons at frequency \(\omega\): 
\begin{eqnarray}
w_{1n}=\frac{1}{T_{n1}}, w_{n1}=\frac{r_n^T}{T_{n1}}, w_{23}=\frac{1}{T_{32}}, w_{32}=\frac{1}{T_{32}}\frac{r_3^T}{r_2^T}.\nonumber\\
\label{rates}
\end{eqnarray}
Then we get from Eq.~\formref{rho1},\formref{sigma2},\formref{rates} the following expressions for the stationary populations:
\begin{widetext}
\begin{equation}
\frac{\rho_{nn}}{\rho_{11}}=\frac{r_n^T\brac{\frac{1}{T_{21}T_{31}}+\frac{1}{T_{21}T_{32}}+\frac{1}{T_{31}T_{32}}\frac{r_3^T}{r_2^T}}+
\frac{2\abs{\Omega}^2}{\gamma_{32}}\brac{\frac{r_2^T}{T_{21}}+\frac{r_3^T}{T_{31}}}}
{\brac{\frac{1}{T_{21}T_{31}}+\frac{1}{T_{21}T_{32}}+\frac{1}{T_{31}T_{32}}\frac{r_3^T}{r_2^T}}+
\frac{2\abs{\Omega}^2}{\gamma_{32}}\brac{\frac{1}{T_{21}}+\frac{1}{T_{31}}}},  n=2,3
\label{popul}
\end{equation}
\end{widetext}
Consider the condition of EIT Eq.~\formref{EITcond}, that is realized if \(1/T_{21}\ll 1/T_{31},1/T_{32}\) and if the number of thermal photons at high frequencies is low \(T\ll \hbar\omega_{31}\). Meanwhile the number of thermal photons at splitting frequency \(\omega_{21}\) may be high. It corresponds to real experimental conditions.  Thus, for example, for Rb vapor the number of thermal photons at splitting frequency \(\omega_{21}=2\pi\times 6.83 GHz\) becomes of the order of unity at temperature \(T\sim 0.5 K\), while at probe frequencies (\(\lambda=794nm\)) the number of thermal photons is much less than unity even for the room temperature. We put the exact equality, that corresponds to reality with good accuracy:
\begin{eqnarray}
r_3^T=0.\nonumber
\end{eqnarray}

It is remarkable that according to Eq.~\formref{popul} under EIT condition the level \(\cet{2}\) is devastated  due to the action of strong drive wave and fast relaxation from the upper level, that is manifestation of well known CPT effect:
\begin{eqnarray}
\frac{\rho_{22}}{\rho_{11}}\approx r_2^T\brac{1+\frac{T_{31}}{T_{32}}}\frac{\gamma_{32}}{2T_{21}\abs{\Omega}^2}\ll r_2^T.
\label{devastation}
\end{eqnarray}
Nevertheless the residual population of the level \(\cet{2}\) will be taken into account. And, as it will be shown later, it is necessary to do for correct noise estimation. 

Note, that the upper level is slightly populated due to drive field action (although the number of atoms at upper level does not depend on drive intensity under EIT condition):
\begin{eqnarray}
\frac{\rho_{33}}{\rho_{11}}\approx r_2^T\frac{T_{31}}{T_{21}}\ll\frac{\rho_{22}}{\rho_{11}}.
\label{rho_33}
\end{eqnarray}
but this population can be ignored.

\subsection{Hot medium susceptibility}

The amplitudes of slowly varying components of polarization \(\hat{\mathbf{P}}=\sum_{j=d, \alpha, \beta}\brac{\hat{\mathbf{P}}_{j}\brac{z,t}e^{i{k_j}z-i\omega_j t}
+\hat{\mathbf{P}}_{j}^\dag\brac{z,t}e^{-i{k_j}z+i\omega_j t}}\) are expressed via density components in the following way:
\begin{eqnarray}
\hat{\mathbf{P}}_{j}=\mathbf{d}_{23}\hat{\sigma}_{32}^j+\mathbf{d}_{13}\hat{\sigma}_{31}^j.
\label{P_sigma}
\end{eqnarray}

We solve the dynamical part of equations for the atomic coherences Eqs.~\formref{sigma1},
taking into account small population of the level \(\cet{2}\) (the population of level \(\cet{3}\) can be ignored as dictated by Eq.~\formref{rho_33}).  As a result,  we calculate the susceptibility components \(\chi_{\alpha}^{H, aH}\), \(\chi_{\beta}^{H, aH}\), \(\chi_{\alpha\beta}\), \(\chi_{\beta\alpha}\) defined by relations:
\begin{eqnarray}
\hat{P}_{\alpha_\nu}=\chi^H_{\alpha}(\nu)E_\alpha\hat{c}_{\alpha_\nu}+\chi^{aH}_{\alpha}(\nu)E_\alpha\hat{c}_{\alpha_\nu}+\nonumber\\
+\chi_{\alpha\beta}(\nu)e^{i2\theta}E_\beta\brac{\hat{c}_{\beta_{-\nu}}}^\dag
+\delta P_{\alpha_\nu}^L \nonumber\\
\hat{P}_{\beta_\nu}=\chi^H_{\beta}(\nu)E_\beta\hat{c}_{\beta_\nu}+\chi^{aH}_{\beta}(\nu)E_\beta\hat{c}_{\beta_\nu}+\nonumber\\
+\chi_{\beta\alpha}(\nu)e^{i2\theta}E_\alpha\brac{\hat{c}_{\alpha_{-\nu}}}^\dag
+\delta P_{\beta_\nu}^L. \nonumber\\
\label{P1}
\end{eqnarray}
The following expressions are obtained:
\begin{eqnarray}
\chi^H_{\alpha}(\nu)=\frac{\eta}{4\pi\abs{\Omega}}\brac{\frac{\omega_{\alpha}+\nu-\omega_{31}}{\abs{\Omega}}-\abs{\zeta}^2\frac{\abs{\Omega}}{\omega_{21}}};\nonumber\\
\chi^{aH}_\alpha(\nu)=\nonumber\\
\frac{i\eta}{4\pi\abs{\Omega}}\brac{\frac{\brac{\omega_{\alpha}+\nu-\omega_{31}}^2\gamma_{31}}{\abs{\Omega}^3}+\frac{\gamma_{21}}{\abs{\Omega}}-\frac{\abs{\Omega}}{\gamma_{32}}\frac{n_{23}}{\rho_{11}}}\approx;\nonumber\\
\approx
\frac{i\eta}{4\pi\abs{\Omega}}\brac{\frac{\nu^2\gamma_{31}}{\abs{\Omega}^3}+\frac{\gamma_{21}}{\abs{\Omega}}-\frac{\abs{\Omega}}{\gamma_{32}}\frac{n_{23}}{\rho_{11}}};\nonumber\\
\chi^H_\beta(\nu)=\frac{\eta}{8\pi\omega_{21}}\brac{1-\frac{3}{2}\frac{\omega_\beta+\nu-2\omega_{32}+\omega_{31}}{\omega_{21}}};\nonumber\\
\chi^{aH}_\beta(\nu)=\frac{i\eta}{8\pi\omega_{21}}\brac{-\frac{3}{2}\frac{\gamma_{31}}{\omega_{21}}+
2\frac{1}{\omega_{21}}\brac{\gamma_{31}+\frac{\gamma_{32}}{\abs{\zeta}^2}}\frac{n_{23}}{\rho_{11}}};\nonumber\\
\chi_{\alpha\beta}(\nu)=\chi_{\beta\alpha}(\nu)=-\frac{\eta}{4\pi\omega_{21}}.\nonumber\\
\label{chi_ps}
\end{eqnarray}
Here  
\begin{eqnarray}
\eta=4\pi\abs{d_{31}}^2N/\hbar, 
\label{eta}
\end{eqnarray}
\(N\) is the density of atoms. The relations Eq.~\formref{chi_ps} correctly describe the frequency dependence of the susceptibility components in a central part of EIT window, more precisely for \(\abs{\nu}<\abs{\Omega}\sqrt{\gamma_{21}/\gamma_{31}}\). The small corrections of the order \(\gamma_{31}\abs{\Omega}^2/\brac{\gamma_{21}\omega_{21}^2}\ll1\) (Eq.~\formref{small}) are not taken into account. In particular it means that the frequency dependence of anti-Hermitian part of \(\alpha\)-wave susceptibility is essential only at the scale larger than the frequency band of parametric instability \(\abs{\nu}\gtrsim \abs{\Omega}^2/\omega_{21}\) (see Sec.~\ref{key parameters}).

We use the obtained expression for the population of level \(\cet{2}\) Eq.~\formref{devastation} 
and the relation between longitudinal and transverse relaxation rates:
\begin{eqnarray}
\gamma_{21}=\frac{1}{2}A_{21}\brac{2n_T(\omega_{21})+1}+\Gamma_{21}=\gamma_{21}^0+A_{21}n_T(\omega_{21}),\nonumber\\
\gamma_{21}^0=\frac{1}{2}A_{21}+\Gamma_{21},\nonumber\\
\label{dephasing rate}
\end{eqnarray}
where \(\Gamma_{21}\) is dephasing rate at low-frequency transition \(\cet{2}-\cet{1}\) caused by elastic processes.
Then we can obtain the explicit temperature dependence of anti-Hermitian part of partial susceptibility of \(\alpha\)-wave: 
\begin{eqnarray}
\chi^{aH}_\alpha=\frac{i\eta}{4\pi\abs{\Omega}}\brac{\frac{\nu^2\gamma_{31}}{\abs{\Omega}^3}+\frac{\gamma_{21}^0}{\abs{\Omega}}+\frac{A_{21}n_T(\omega_{21})}{2\abs{\Omega}}\brac{1-\frac{A_{32}}{A_{31}}}}.\nonumber\\
\label{chiT}
\end{eqnarray}
It follows from Eq.~\formref{chiT} that the linear decrement of \(\alpha\)-wave is modified in the presence of thermal excitations only under condition of essentially different Einstein coefficients at two high frequency transitions: \(\cet{3}-\cet{1}\) and \(\cet{3}-\cet{2}\). In a narrow range of parameters it can even became negative, that corresponds to so called Amplification Without Inversion regime \cite{Kochar_Mandel_Radion}. Considering the case of close relaxation rates at high frequencies \(A_{31}\approx A_{32}\) we get, that anti-Hermitian part of partial susceptibility of \(\alpha\)-wave is not sensitive to the thermal excitations \(\chi^{aH}_\alpha\approx\chi^{aH}_\alpha(n_T(\omega_{21})=0)\). Note that the implicit temperature dependence of linear decrement of \(\alpha\)-wave is defined by temperature dependence of dephasing rate \(\Gamma_{21}(T)\).

It is interesting to note that the medium driven by resonant coherent field has the negative conductivity at frequency \(\omega_\beta\): \(-i\chi^{aH}_{\beta}
<0\). The negative dissipation is caused by the  process of Raman scattering of the drive wave at the transition \(\cet{1}-\cet{2}\) into Stokes satellite at frequency \(\omega_\beta\).
The absolute value of this amplification is much less than the absorption at the frequency of \(\alpha\)-wave due to condition Eq.~\formref{small}:
\begin{eqnarray}
\abs{\chi_\beta^{aH}}\ll\abs{\chi_\alpha^{aH}}.
\label{beta_alpha}
\end{eqnarray}
The dependence of anti-Hermitian component of susceptibility at \(\omega_\beta\) on the thermal redistribution among atomic levels is presented in Eq~\formref{chi_ps}, but it is unessential due to Eq.~\formref{devastation}. The analogous dependences in the Hermitian parts of susceptibilities are omitted.

So we can conclude that due to field induced devastation of level \(\cet{2}\) (Eq.~\formref{devastation}) under EIT conditions Eq.~\formref{EITcond} all components of medium susceptibility can be considered not sensitive to the thermal excitations of atoms in a wide range of parameters. 

The radical dependence on temperature in the EIT medium appears in the noise terms of polarizations Eq.~\formref{P1}, namely in their correlation functions.

\subsection{Correlation functions for the noise components of the medium polarization \label{noise}}

The commutators of noise polarization components, defined in correspondence with Eq.~\formref{commutatorLnu3} by the linear partial decrements of waves, do not depend on the temperature, as follows from conclusion of previous section. The situation is different with correlators separately or their sum. 

In linear medium the relation between correlation functions and the linear decrement is dictated by standard Fluctuation Dissipation Theorem \cite{Callen}:
\begin{eqnarray}
\bracm{\delta\hat{P}_{j\nu}^{L}\brac{z}\delta\hat{P}^{L\dag}_{j\nu'}\brac{z'}+\delta\hat{P}^{L\dag}_{j\nu'}\brac{z'}\delta\hat{P}^{L}_{j\nu}\brac{z}}
=\nonumber\\-i\frac{\hbar}{\pi}\frac{1}{S_\perp}\chi^{aH}_{j}(\nu)\brac{2n_T\brac{\omega_j+\nu}+1}\delta\brac{\nu-\nu'}\delta\brac{z-z'}.\nonumber\\
\label{FDT}
\end{eqnarray}   

We apply the technique developed in paper \cite{ErukhOL15} to calculate the correlation functions of the noise components of polarisation excited in different frequency intervals, corresponding to \(\alpha\)- and \(\beta\)-waves, in FWM regime beyond simple RWA (but using condition Eq.~\formref{small}). Namely, we find the noise solution of Eqs.~\formref{sigma1}, and using correlation functions for the atomic noise operators Eqs.~\formref{mnnm_omega}, \formref{mabm_omega}, we get for correlation functions of noise components of polarizations Eq.~\formref{P_sigma} the following expressions:
\begin{eqnarray}
\bracm{\delta\hat{P}_{\alpha\nu}^{L}\brac{z}\delta\hat{P}^{L\dag}_{\alpha\nu'}\brac{z'}\mp\delta\hat{P}^{L\dag}_{\alpha\nu'}\brac{z'}\delta\hat{P}^{L}_{\alpha\nu}\brac{z}}=\nonumber\\
-i\frac{\hbar}{\pi}\frac{1}{S_\perp}\frac{i\eta}{4\pi \abs{\Omega}}\brac{\frac{\nu^2\gamma_{31}}{\abs{\Omega}^3}+\frac{\gamma_{21}}{\abs{\Omega}}\mp\frac{\abs{\Omega}}{\gamma_{32}}\frac{n_{23}}{\rho_{11}}}\times\nonumber\\
\times\delta(\nu-\nu')\delta(z-z')\nonumber\\
\label{corr1}
\end{eqnarray}
\begin{eqnarray}
\bracm{\delta\hat{P}_{\beta\nu}^{L}\brac{z}\delta\hat{P}^{L\dag}_{\beta\nu'}\brac{z'}\mp\delta\hat{P}^{L\dag}_{\beta\nu'}\brac{z'}\delta\hat{P}^{L}_{\beta\nu}\brac{z}}=\nonumber\\
-i\frac{\hbar}{\pi}\frac{1}{S_\perp}\frac{i\eta}{8\pi\omega_{21}}\times\nonumber\\
\times\brac{\frac{\gamma_{31}}{2\omega_{21}}\mp \frac{2\gamma_{31}}{\omega_{21}}
+\frac{2}{\omega_{21}}\brac{\gamma_{31}+\frac{\gamma_{32}}{\abs{\zeta}^2}}\frac{n_{23}}{\rho_{11}}}\times\nonumber\\
\times\delta(\nu-\nu')\delta(z-z')\nonumber\\
\label{corr2}
\end{eqnarray}
\begin{eqnarray}
\bracm{\delta\hat{P}_{\alpha\nu}^{L}\brac{z}\delta\hat{P}^{L\dag}_{\beta\nu'}\brac{z'}\mp\delta\hat{P}^{L\dag}_{\beta\nu'}\brac{z'}\delta\hat{P}^{L}_{\alpha\nu}\brac{z}}=0\nonumber\\
\label{corr3}
\end{eqnarray}
\begin{eqnarray}
\bracm{\delta\hat{P}_{\alpha\nu}^{L}\brac{z}\delta\hat{P}^{L}_{\beta\nu'}\brac{z'}\mp
\delta\hat{P}^{L}_{\beta\nu'}\brac{z'}\delta\hat{P}^{L}_{\alpha\nu}\brac{z}}=\nonumber\\
-i\frac{\hbar}{\pi}\frac{1}{S_\perp}\frac{i\eta e^{2i\theta}}{8\pi\omega_{21}}\times\nonumber\\
\times\brac{\frac{2\gamma_{31}\brac{\omega_{31}-\omega_p-\nu}}{\abs{\Omega}^2}+\frac{2\gamma_{31}}{\omega_{21}}\mp\frac{2in_{23}}{\rho_{11}}}\times\nonumber\\
\times\delta(\nu+\nu')\delta(z-z')\nonumber\\
\label{cross}
\end{eqnarray}
First that we can see, the obtained commutators for the noise polarizations are exactly coincide with the corresponding relation Eq.~\formref{commutatorLnu3}, that guarantees the fulfilment of commutation relations for the field operators:
\begin{eqnarray}
\bracr{\delta\hat{P}_{j\nu}^{L}\brac{z}\delta\hat{P}^{L\dag}_{j\nu'}\brac{z'}}=-i\frac{\hbar}{\pi}\frac{1}{S_\perp}\chi^{aH}_{j}(\nu)\delta\brac{\nu-\nu'}\delta\brac{z-z'},\nonumber\\
\label{commutator}
\end{eqnarray}
where \(\chi^{aH}_{j}(\nu)\) where obtained in Eq.~\formref{chi_ps}.
Second, the sums of correlation functions can be written in form that resemble the standard FDT relation Eq~\formref{FDT}, so that they are proportional to the anti-Hermitian components for \(\alpha\)- and \(\beta\)-waves, but with cardinally different proportionality factor:
\begin{eqnarray}
\bracm{\delta\hat{P}_{j\nu}^{L}\brac{z}\delta\hat{P}^{L\dag}_{j\nu'}\brac{z'}+\delta\hat{P}^{L\dag}_{j\nu'}\brac{z'}\delta\hat{P}^{L}_{j\nu}\brac{z}}
=\nonumber\\=\abs{-i\frac{\hbar}{\pi}\frac{1}{S_\perp}\chi^{aH}_{j}(\nu)}\brac{2S_j(\nu)+1}\delta\brac{\nu-\nu'}\delta\brac{z-z'}.\nonumber\\
\label{FDT_new}
\end{eqnarray}
The coefficients \(S_j\) are related to the polarization noise associated to the processes of spontaneous emission, they define the excess noise beyond that which is necessary to preserve the canonical commutation relation for the field operators.  

For the \(\alpha\)-wave this coefficient is equal to:
\begin{eqnarray}
S_\alpha(\nu)=\frac{\frac{\abs{\Omega}^2}{\gamma_{32}\gamma_{21}}\frac{n_{23}}{\rho_{11}}}{1-\frac{\abs{\Omega}^2}{\gamma_{32}\gamma_{21}}\frac{n_{23}}{\rho_{11}}+\frac{\nu^2\gamma_{31}}{{\abs{\Omega}}^2\gamma_{21}}
}
=\frac{A_{21}}{\Gamma_{21}}\frac{n_T(\omega_{21})}{1+\frac{A_{21}}{2\Gamma_{21}}+\frac{\nu^2\gamma_{31}}{\abs{\Omega}^2 \Gamma_{21}}}.\nonumber\\
\label{S_p}
\end{eqnarray}
Note that it is proportional to the population of the devastated level \(\cet{2}\), but with large coefficient of proportionality \(\frac{\abs{\Omega}^2}{\gamma_{32}\gamma_{21}}\). So that using relation Eq.~\formref{devastation} for the population \(\rho_{22}\), relation Eq.~\formref{dephasing rate} for the low-frequency coherence decay rate \(\gamma_{21}\) we get the last expression. (It is written under assumption \(A_{31}\approx A_{32}\).) 
It is important, that the thermal noise of polarization at the frequency \(\omega_\alpha\) is defined by the averaged number of thermal photons at low frequency of ground-state splitting \(n_T(\omega_{21})\), and depends on the ratio of spontaneous emission at low frequency transition  \(A_{21}\) to the dephasing rate caused by elastic processes \(\Gamma_{21}\).  It is maximal at zero detuning \(\nu\) and it is the stronger the larger ratio of population exchange  rate to the rate of low frequency coherence relaxation arisen from elastic collisions or atoms moving in a and out of the interaction region. The dependence of noise at frequency \(\omega_\alpha\) on the parameters of transition with frequency \(\omega_{21}\) is explained by the process of parametric transfer of noise from low to high frequency in the presence of resonant strong drive wave. It can be also interpreted as result of spontaneous anti-Stokes Raman scattering, but modified under conditions of resonant interaction.     

The analogous coefficient for the \(\beta\)-wave: 
\begin{equation}
S_\beta(\nu)=1/3.
\label{S_s}
\end{equation} 
It is interesting, that it is not defined by nonzero excitations in the atomic system (thermal distribution among levels). The reason is that the fluctuations at frequency \(\omega_\beta\) with nonzero flux can be generated due to scattering of the drive wave on zero fluctuations of reservoir at low frequency.  In other words it is caused by the spontaneous Stokes Raman scattering. 
For the following analysis of noise characteristics of generated two-mode squeezed radiation it was important here to find out that this coefficient is not large, so that the noise source at the frequency \(\omega_\beta\) can be ignored, since the power spectral density of noise source at the frequency \(\omega_\beta\) is proportional to the anti-Hermitian part of susceptibility at the frequency \(\omega_\beta\) and it is much smaller than the the noise source at frequency \(\omega_\alpha\) because Eq.~\formref{beta_alpha} is fulfilled. The same concerns the cross-correlation terms Eqs.~\formref{cross}.

\subsection{The solution of the coupled equations for the field operators}

Using the obtained relations for the partial and the parametric components of susceptibilities Eq.~\formref{chi_ps} we can derive the coupled equations for the filed operators according to Eq.~\formref{prop2}. Note that these equations were derived under condition that the nonlinear and dissipative components of polarization are small compared with the field. It is reduced to the condition imposed on the medium density:
\begin{equation}
\frac{\eta}{\omega_{21}}\ll 1,
\label{eta_sm}
\end{equation}
where \(\eta\) is given by Eq.~\formref{eta}.
So we get the following coupled equations for opposite (\(\nu\) and \(-\nu\)) spectral components of the field operators corresponding for \(\alpha\) and \(\beta\)-waves: 
\begin{eqnarray}
\frac{\partial \hat{s}_{\alpha_\nu}\brac{z}}{\partial z}-\frac{i\nu}{v_{gr\alpha}}\hat{s}_{\alpha_\nu}\brac{z}+\kappa_\alpha\hat{s}_{\alpha_\nu}\brac{z}=\nonumber\\
=-i\chi e^{2i\theta}\hat{s}_{\beta_{-\nu}}^\dag+\hat{L}_{\alpha_\nu},\nonumber\\
\frac{\partial \hat{s}_{\beta_{-\nu}}^\dag\brac{z}}{\partial z}-\frac{i\nu}{v_{gr\beta}}\hat{s}_{\beta_{-\nu}}^\dag\brac{z}+\kappa_\beta\hat{s}_{\beta_{-\nu}}^\dag\brac{z}=\nonumber\\
=i\chi e^{-2i\theta}\hat{s}_{\alpha_{\nu}}+\hat{L}_{\beta_{-\nu}}^\dag\nonumber\\
\label{system}
\end{eqnarray}
Here \(\hat{s}_{j_\nu}=\sqrt{\abs{v_{grj}}}\hat{c}_{j_\nu}\), as defined in Sec.~\ref{The field equations}, are the operators that define the energy fluxes along the \(z\)-axis in \(\alpha\) and \(\beta\) waves. The group velocities along the \(z\)-axis are defined by partial susceptibilities \(\chi_{j}^{H}\) (Eq.~\formref{chi_ps}) in such a way: \(v_{grj}=c/\brac{n_0+2\pi\frac{\omega_j} {n_0}\frac{\partial\chi_j^H}{\partial\nu}}\), where \(n_0\) is the refractive index of the background. So we get:
\begin{eqnarray} 
v_{gr\alpha}\approx \frac{2cn_0\abs{\Omega}^2}{\eta\omega_\alpha}\nonumber\\
v_{gr\beta}\approx \frac{c}{n_0-\frac{3\eta\omega_\beta}{8n_0\omega_{21}^2}}.\nonumber
\end{eqnarray}
The expression for the \(\alpha\)-wave group velocity is presented under condition of strong slowdown:
\begin{equation}
\frac{\eta\omega_{\alpha}}{\abs{\Omega}^2}\gg 1.
\label{eta_large_1}
\end{equation}
It is interesting to note that under condition \(\abs{1-3\eta\omega_\beta/8n_0^2\omega_{21}^2}<n_0^{-1}\) the group velocity of \(\beta\)-wave exceeds the light velocity, and under condition 
\begin{equation}
\frac{3\eta\omega_{\beta}}{8n_0^2\omega_{21}^2}> 1
\label{eta_large_2}
\end{equation}
it changes sign to negative. Such effects are well known for active and dissipative media and do not lead to violation of basic principles (see for example \cite{Brillouin, Wang, Kryachko}). 
Note that both conditions Eqs.~\formref{eta_large_1},\formref{eta_large_2} are compatible with condition Eq.~\formref{eta_sm}. 

The absorption coefficient of \(\alpha\)-wave and amplification coefficient for \(\beta\)-wave are equal 
\begin{eqnarray}
\kappa_{\alpha}=-\frac{2\pi\omega_{\alpha}}{cn_{0}}i\chi^{aH}_{\alpha}\approx \frac{\eta\omega_\alpha}{2cn_0\abs{\Omega}}\brac{\frac{\nu^2\gamma_{31}}{\abs{\Omega}^3}+\frac{\gamma_{21}^0}{\abs{\Omega}}},\nonumber\\ 
-\kappa_{\beta}=\frac{2\pi\omega_{\beta}}{cn_0}i\chi^{aH}_{\beta}\approx\frac{\eta\omega_\beta}{2cn_0\omega_{21}}\frac{3\gamma_{31}}{4\omega_{21}}.\nonumber\\
\label{kappa_p}
\end{eqnarray}
The Langevin noise operators: \(\hat{L}_{\alpha,\beta_\nu}=i\sqrt{\frac{2\pi\omega_{\alpha,\beta}}{\hbar cn_0}}\delta\hat{P}^L_{\alpha,\beta_\nu}\). The coefficient of parametric coupling in Eq.~\formref{system} is equal to
\begin{eqnarray}
\chi=\frac{\eta\sqrt{\omega_\alpha\omega_\beta}}{2cn_0\omega_{21}}\approx\frac{\eta\omega_\alpha}{2cn_0\omega_{21}}.
\label{chi}
\end{eqnarray}

The solution of the system of equations Eq.~\formref{system} takes the form:
\begin{eqnarray}
\begin{pmatrix} \hat{s}_{\alpha_\nu}(z)\\ \hat{s}_{\beta_{-\nu}}^\dag(z) \end{pmatrix}=\nonumber\\
\begin{pmatrix} 1\\ K_{X} \end{pmatrix}e^{iq_X(\nu)z}\brac{\hat{u}_{X_\nu}+\int_0^z{e^{-iq_X(\nu)\xi}\hat{f}_{X_\nu}}d\xi}+\nonumber\\
+\begin{pmatrix} 1\\ K_{O} \end{pmatrix}e^{iq_O(\nu)z}\brac{\hat{u}_{O_\nu}+\int_0^z{e^{-iq_O(\nu)\xi}\hat{f}_{O_\nu}}d\xi},\nonumber\\
\label{solut1}
\end{eqnarray}
where
\begin{eqnarray}
q_{O,X}=\frac{\nu}{2}\brac{\frac{1}{v_{gr\alpha}}+\frac{1}{v_{gr\beta}}}+i\frac{\kappa_\alpha-\abs{\kappa_\beta}}{2}\pm \chi\sqrt{\sigma^2-1},\nonumber\\
\sigma=\frac{\nu}{2\chi}\brac{\frac{1}{v_{gr\alpha}}-\frac{1}{v_{gr\beta}}}+\frac{i}{2\chi}\brac{\kappa_\alpha+\abs{\kappa_\beta}},\nonumber\\
K_{O,X}=\brac{\sigma\mp\sqrt{\sigma^2-1}}e^{-2i\theta}.\nonumber\\
\label{solut2}
\end{eqnarray}
As in  papers \cite{Tokman, Vdovin}, we denote two normal modes as the O- and X-mode. The coefficients \(K_{O}\) and \(K_{X}\) define the ratio between waves with frequencies \(\omega_{\alpha,\beta}\) in normal modes. The amplitudes \(\hat{u}_{X_\nu}\) and \(\hat{u}_{O_\nu}\) are expressed via the operators of incident radiation \(\hat{c}_{{\alpha,\beta}_\nu}(0)\) defined in section \(z=0\) in vacuum using boundary condition Eq.~\formref{boundary}:
\begin{eqnarray}
\hat{u}_{X_\nu}=\sqrt{c}\frac{K_{O}\hat{c}_{\alpha_\nu}(0)-\hat{c}^\dag_{\beta_{-\nu}}(0)}{K_O-K_X}\nonumber\\
\hat{u}_{O_\nu}=\sqrt{c}\frac{\hat{c}^\dag_{\beta_{-\nu}}(0)-K_{X}\hat{c}_{\alpha_\nu}(0)}{K_O-K_X}.
\label{solut3}
\end{eqnarray}
The operators \(\hat{f}_{X_\nu}\) and \(\hat{f}_{O_\nu}\) are the Langevin sources for O and X modes:
\begin{eqnarray}
\hat{f}_{X_\nu}=\frac{K_O\hat{L}_{\alpha_\nu}-\hat{L}^\dag_{\beta_{-\nu}}}{K_O-K_X}\nonumber\\
\hat{f}_{O_\nu}=\frac{\hat{L}^\dag_{\beta_{-\nu}}-K_X\hat{L}_{\alpha_\nu}}{K_O-K_X}.
\label{solut4}
\end{eqnarray}
We take into account strong group deceleration of \(\alpha\)-wave and acceleration of \(\beta\)-wave, so that \(\abs{v_{gr\alpha}/v_{gr\beta}}\ll 1\), and neglect weak partial amplification coefficient of \(\beta\)-wave with respect to partial absorption of \(\alpha\)-wave \(\abs{\kappa_\beta}\ll\kappa_\alpha\) (Eq.~\formref{beta_alpha}). Then the relations Eqs.~\formref{solut2} can be simplified:
\begin{eqnarray}
q_{O,X}=\chi\brac{\sigma\pm\sqrt{\sigma^2-1}},\nonumber\\
\sigma=\frac{\nu}{2\chi v_{gr\alpha}}+\frac{i \kappa_\alpha}{2\chi},\nonumber\\
K_{O,X}=\brac{\sigma\mp\sqrt{\sigma^2-1}}e^{-2i\theta}.\nonumber\\
\label{solut2_1}
\end{eqnarray}

Analysing the dynamical part of solution Eq.~\formref{solut1} we can make the following conclusions.

Under ideal condition of no dissipation \(\kappa_\alpha=0\) the limitation of frequency detuning:
\begin{eqnarray}
\abs{\nu}<\Delta_P,\nonumber
\end{eqnarray}
where \(\Delta_P=2\chi v_{gr\alpha}={2\abs{\Omega}^2}/{\omega_{21}}\) (see Eq.~\formref{Delta_P}),
defines the band of parametric instability. If \(\kappa_\alpha=0\) and  \(\nu=0\) two ideally correlated waves are generated with equal amplitudes, so that we have:
\begin{eqnarray}
q_{O,X}=\pm i\chi,\nonumber\\
K_{O,X}=\mp ie^{-2i\theta}.
\label{solut2_2}
\end{eqnarray}

Under condition of nonzero dissipation \(\kappa_\alpha\neq 0\) X-mode is always amplified (at every detuning) (see \cite{Tokman, ErukhPRA11}), but the ratio between partial amplitudes in normal modes is changed, so that the correlations are spoiled. The regime of strong parametric instability when the solution is close to ideal Eq.~\formref{solut2_2} is realized if the following conditions are fulfilled (see Eq.~\formref{strong}):
\begin{eqnarray}
\frac{\chi}{\kappa_\alpha}=\frac{\abs{\Omega}^2}{\gamma_{21}^0\omega_{21}}\gg 1
\label{weak_diss}
\end{eqnarray}
and 
\begin{eqnarray}
\abs{\nu}\ll\Delta_P.\nonumber
\end{eqnarray}

\section{\label{squeezing} Two-mode squeezing}
In the considered regime of parametric instability of bichromatic radiation the flow of correlated photons is generated at the output of interaction section, and the generated radiation exhibits characteristics of two-mode squeezing.  So that the fluctuations of sum of one quadratures of the field components and difference of the other quadratures falls below that of the vacuum state \cite{Lvovsky}. We'll examine the fluctuation of the following observable: 
\begin{eqnarray}
\hat{X}_{\nu}=\frac{1}{\sqrt{2}}\brac{\hat{X}_{\alpha_{\nu}}-\hat{X}_{\beta_{-\nu}}},\nonumber
\end{eqnarray}
where
\begin{eqnarray}
\hat{X}_{j_\nu}=\hat{c}_{j_\nu}e^{i\varphi_j}+\brac{\hat{c}_{j_\nu}}^\dag e^{-i\varphi_j}\nonumber\\
j=\alpha,\beta.\nonumber
\end{eqnarray}
Here the operators \(\hat{c}_{j_\nu}\) are defined by the ''flux`` amplitudes Eqs.~\formref{solut1} at the output of the layer \(\hat{c}_{j_\nu}=\frac{1}{\sqrt{c}}\hat{s}_{j_\nu}(l)\).
The  observable \(\hat{X}_{\nu}\) and its fluctuation may correspond to the measured spectral characteristics of current at the detector in balanced homodyne detection scheme,
where phases \(\varphi_{\alpha,\beta}\) are determined by the parameters of local oscillator \cite{Yuen}.   
More precisely, we'll calculate the normalized spectral noise power defined by the averaged quadratic fluctuations of the observable \(\hat{X}_{\nu}\) as it is related to the standard quantum limit (SQL) value. Define this quantity as
\begin{eqnarray}
Q_\nu=\brac{2\pi S_\perp c\Delta\nu_{det}}\overline{\bracm{\brac{\Delta\hat{X}_{\nu}}^2}},
\label{Pnoise}
\end{eqnarray}
where the overline corresponds to the averaging over detection resolution bandwidth \(\Delta\nu_{det}\).
Here we take into account that \(\overline{\bracm{\brac{\Delta\hat{X}_{\nu}}^2}}_{SQL}=1/(2\pi S_\perp c\Delta\nu_{det})\).

\subsection{The general solution for two-mode fluctuations}

Consider the boundary condition at \(z=0\) corresponding to a completely uncorrelated state of vacuum fluctuations, so that \(\bracm{\hat{c}_{i\nu}(0)\hat{c}_{j\nu'}^{\dag}(0)}=\frac{\delta(\nu-\nu')\delta_{ij}}{2\pi S_{\perp}c}\). With aim to calculate the quantity \(P_{\nu}^{noise}\) we use the obtained solution of equation for parametrically coupled waves Eqs.~\formref{solut1},\formref{solut3},\formref{solut4} with coefficients Eq.~\formref{solut2_1} and the expressions for the correlation functions of the noise components of the polarizations Eqs.~\formref{corr1},\formref{corr2},\formref{corr3},\formref{cross}. 

As we have shown previously in Sec.~\ref{noise} the noise spectral density of polarization at the nonresonant frequency \(\omega_\beta\) is small compared to the of that at the resonant frequency \(\omega_\alpha\) because we stay within the framework of negligible dissipation of \(\beta\)-wave in comparison with dissipation of \(\alpha\)-wave Eq.~\formref{small}, and the noise associated with Raman spontaneous emission processes can not significantly increase the level of fluctuations at frequency \(\omega_\beta\) (see Eq.~\formref{FDT_new},\formref{S_s}). 
The analysis of general solution for the fluctuation of joint quadrature operator \(\hat{X}_{\nu}\) shows that taking into account nonzero noise spectral density of polarization at the  frequency \(\omega_\beta\) as well as cross-spectral density between noise polarizations at  frequencies \(\omega_\beta\) and \(\omega_\alpha\) goes beyond the approximation Eq.~\formref{small}. 
So here we introduce the relation for the spectral noise power of the joint observable  \(\hat{X}_{\nu}\) under the assumption that the only source of fluctuations is the quantum and thermal noise associated with the \(\alpha\)-field losses:
\begin{eqnarray}
Q_\nu=\nonumber\\
\frac{1}{\abs{\tilde{K}_0^2-1}^2}
\brac{\frac{1+\abs{K_0}^2}{2}-\brac{\tilde{K}_0\frac{e^{i\Psi}}{2}+\tilde{K_0}^*\frac{e^{-i\Psi}}{2}}}\times\nonumber\\
\brac{1+\abs{K_0}^2-\abs{K_0}^2\frac{\kappa_\alpha\brac{2S_\alpha+1}}{\mathrm{Im} q_X}}e^{-2\mathrm{Im} q_X l}+\nonumber\\
+\frac{1}{\abs{\tilde{K}_0^2-1}^2}
\brac{\frac{1+\abs{K_0}^2}{2}-\brac{\tilde{K}_0^*\frac{e^{i\Psi}}{2}+\tilde{K_0}\frac{e^{-i\Psi}}{2}}}\times\nonumber\\
\brac{1+\abs{K_0}^2+\abs{K_0}^2\frac{\kappa_\alpha\brac{2S_\alpha+1}}{\mathrm{Im} q_X}}e^{-2\mathrm{Im} q_0 l}+\nonumber\\
+\frac{4}{\abs{\tilde{K}_0^2-1}^2}
Re\left(\brac{-\mathrm{Re} \tilde{K_0}+\abs{K_0}^2\frac{e^{i\Psi}}{2}+\frac{e^{-i\Psi}}{2}}\times\right.\nonumber\\
\brac{\mathrm{Re} \tilde{K_0}-i\frac{\abs{K_0}^2}{\abs{K_0}^2-1}\frac{\kappa_\alpha\brac{2S_\alpha+1}}{\chi}}\times
\left. e^{i\brac{q_X-q_0^*}l}\right)\nonumber\\
\label{full}
\end{eqnarray} 
Here \(\tilde{K}_{0,X}=K_{0,X} e^{2i\theta}\), \(\Psi=2\Theta+\varphi_\alpha+\varphi_\beta\). The partial absorption coefficient of \(\alpha\)-wave \(\kappa_\alpha\) is given by relation Eq.~\formref{kappa_p}, the coefficient of parametric coupling \(\chi\) is defined by Eq.~\formref{chi}. The parameter defining the level of thermal fluctuations of polarization at \(\alpha\)-frequency \(S_\alpha\) is given by relation Eq.~\formref{S_p}. In deriving the above relation we have used, that according to
Eq.~\formref{solut2_1} \(\tilde{K}_0\tilde{K}_X=1\), 
\(q_{0,X}=\chi \tilde{K}_{X,0}\), \(\mathrm{Im} q_0=-\mathrm{Im} q_X/\abs{K_0}^2\). 

The resulting relation Eq.~\formref{full} is the sum of three terms. The first one is proportional to the growing exponent with the double increment of X-mode \(\abs{2\mathrm{Im} q_X}\). It corresponds to the exponential growing of noise associated with the instability process. The second one attenuates with the double decrement of the O-mode \(2\mathrm{Im} q_0\). And the third one attenuates slowly with the partial dissipation decrement \(-Re\brac{i\brac{q_X-q_0^*}}=\kappa_\alpha\). The squeezing regime \(Q_\nu\ll 1\) is realized due to appropriate phase \(\Psi\) selection that minimizes the amplitude of the first term, and the most favourable cases if the amplitude of the third term is small, but decrement of O-mode, that determines the squeezing factor, is large.   

The important particular cases following from Eq.~\formref{full} are considered in what follows.

\subsection{No dissipation}
\subsubsection{Center of line}

In ideal case of no dissipation in the medium \(\kappa_p=0\) at the zero detuning \(\nu=0\) when the relations Eq.~\formref{solut2_2} are valid we get:  
\begin{eqnarray}
Q_\nu=\frac{1}{2}\brac{1-\sin\Psi}e^{2\chi l}+\frac{1}{2}\brac{1+\sin\Psi}e^{-2\chi l}.\nonumber
\end{eqnarray}
Choosing \(\Psi=\pi/2\) we get zero amplitude of growing term. Due to equal amplitudes of two waves in the unstable normal mode the exponentially growing fluctuations of \(X_{\alpha,\beta}\)-observables of two waves for chosen phases become ideally correlated and for difference observable these fluctuations will be completely substracted. As result we get:
\begin{eqnarray}
Q_{\nu=0}=e^{-2\chi l}.
\label{center}
\end{eqnarray}
The squeezing improves infinitely with the length of interaction.

\subsubsection{Spectral characteristics}

\begin{figure}
\includegraphics[width=\columnwidth, keepaspectratio=true]{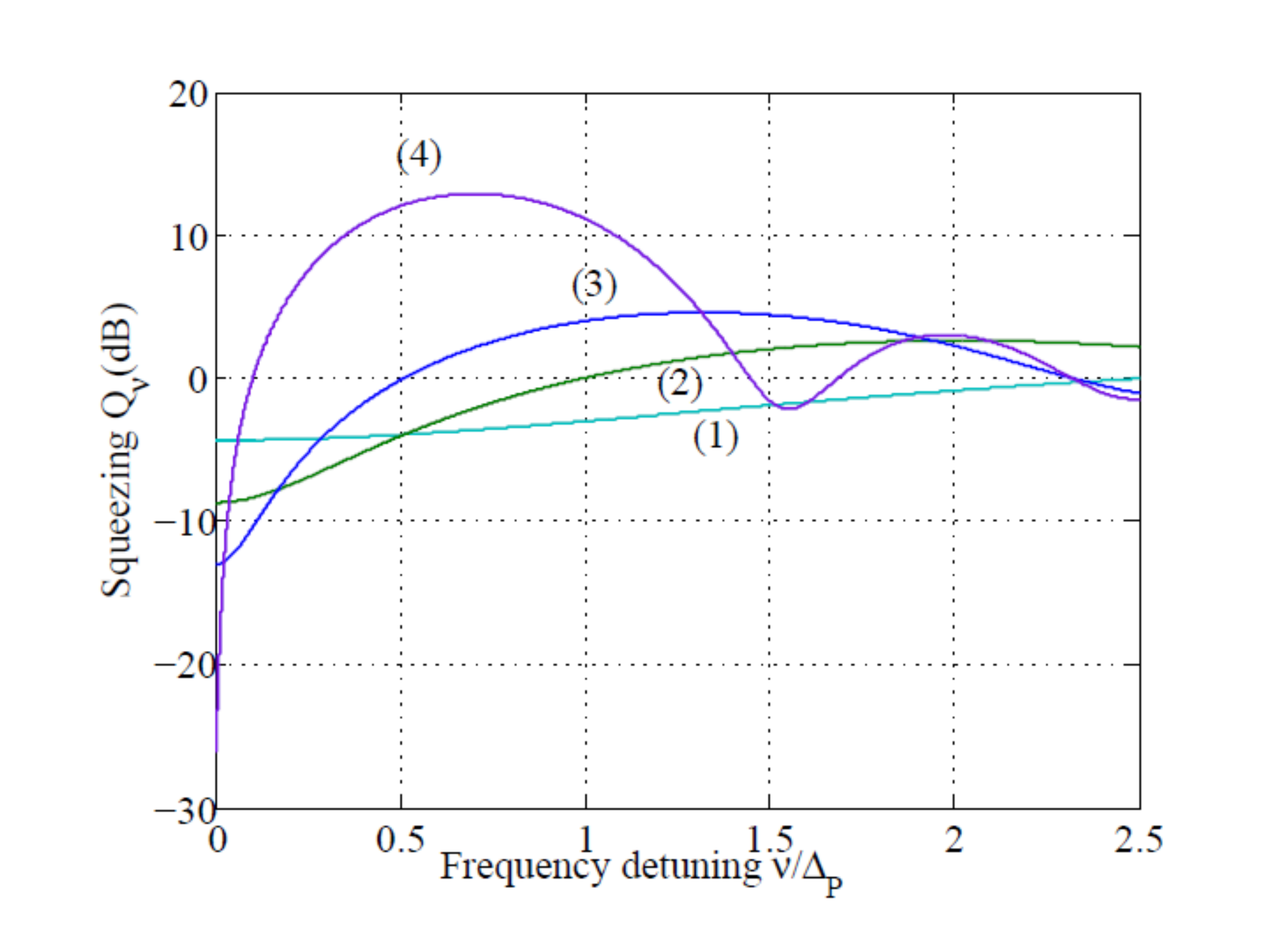}
\caption{\label{figureND} (Color online) Two-mode noise level \(Q_\nu\) (in dB) for the phase \(\Psi=\pi/2\) versus frequency detuning from the FWM resonance \(\nu\) normalized to the frequency band of the parametric instability for the case of negligible dissipation \(\kappa_\alpha=0\), zero temperature \(T=0\), and different lengths of interaction: (1) \(l=0.5/\chi\), (2) \(l=1/\chi\), (3) \(l=1.5/\chi\), (4) \(l=3/\chi\). }
\end{figure}
The noise spectrum under ideal condition of no dissipation (\(\kappa_p=0\)) is given by the following relations obtained from Eq.~\formref{full}. Within the band of parametric instability \(\abs{\nu}<\Delta_P\) we get:
\begin{eqnarray}
Q_\nu = \frac{1}{2\brac{1-\sigma^2}}\brac{1-\cos\brac{\Psi+\delta}}e^{2\chi\sqrt{1-\sigma^2}l}+ \nonumber\\
\frac{1}{2\brac{1-\sigma^2}}\brac{1-\cos\brac{\Psi-\delta}}e^{-2\chi\sqrt{1-\sigma^2}l}+\nonumber\\
 \frac{\sigma}{1-\sigma^2}\brac{\cos\Psi-\cos\delta},\nonumber\\
\label{inband}
\end{eqnarray}
where it is used that  \(\abs{K_0}=1\) and \(\delta\) is defined as \(\tilde{K}_0=e^{i\delta}\).
For the frequency detuning larger than the width of parametric instability band \(\abs{\nu}>\Delta_P\) the noise spectrum is given by:
\begin{eqnarray}
Q_\nu = \frac{\sigma\brac{\sigma-\cos\Psi}}{\sigma^2-1}+\frac{1}{\sigma^2-1}\times\nonumber\\
\mathrm{Re}\brac{\brac{-1+\sigma \cos\Psi-2i\sqrt{\sigma^2-1}\sin\Psi}e^{-2i\chi\sqrt{\sigma^2-1}l}}\nonumber\\
\label{outband}
\end{eqnarray}
In Eqs.~\formref{inband},\formref{outband}  \(\sigma=\nu/\Delta_P\).
It is worth noting  that for every frequency detuning \(\nu_0\) within the band of parametric instability the appropriate phase \(\Psi\) can be found, namely  \(\Psi=-\delta(\nu_0)\),  so that the amplitude of growing term as well as the amplitude of constant term at this frequency become equal to zero and we get
\begin{eqnarray}
Q_{\nu_0}=\frac{1}{1-\sigma^2}e^{-2\chi\sqrt{1-\sigma^2}l}.\nonumber
\end{eqnarray}  
It is obvious that the best squeezing can be achieved in the center of line \(\nu=0\) and the phase \(\Psi=\pi/2\). For this phase the noise level at every detuning \(\nu\neq 0\) decreasing at the initial stage of interaction starts to grow exponentially beginning with the definite length. As a consequence the narrowing of the spectral interval of squeezed light takes place with increase of length of interaction. 
Analysing expressions Eqs.~\formref{inband}, \formref{outband}, we get  the following estimations for  the frequency band of squeezing \(\Delta_{sq}\), defined so that \(Q_\nu<1 (0dB)\) for \(\abs{\nu}<\Delta_{sq}\).  For the short distances \(l\ll 1/\chi\), or more precisely for small density-length product of the medium, the weak squeezing is realized within frequency band 
\begin{eqnarray}
\Delta_{sq}\approx \frac{\pi}{2}\frac{\Delta_P}{\chi l}=\pi\frac{v_{gr\alpha}}{l}.\nonumber
\end{eqnarray}
For \(l=1/{\chi}\) the band of squeezing is equal to the parametric instability band \(\Delta_{sq}=\Delta_P\). And for long distances \(l\gg 1/{\chi}\) (large density-length product)
the band of squeezing becomes much narrower: 
\begin{eqnarray}
\Delta_{sq}\approx 2\Delta_P \exp(-\chi l)\nonumber
\label{width}
\end{eqnarray}
and the magnitude of squeezing in the center becomes much higher in correspondence with Eq.~\formref{center}. The effect of narrowing of the frequency band of squeezed light with the last asymptotic result Eq.~\formref{width} was pointed in paper \cite{Vdovin}, where the process of generation of biband squeezed vacuum in the medium without dissipation was analysed in detail. 

The noise spectrum \(Q_\nu\) in the EIT medium without dissipation \(\kappa_\alpha=0\) is presented for the phase \(\Psi=\pi/2\) at different lengths of interaction at Fig.~\ref{figureND}

\subsection{Dissipative medium. Zero temperature}

\begin{figure}
\includegraphics[width=\columnwidth, keepaspectratio=true]{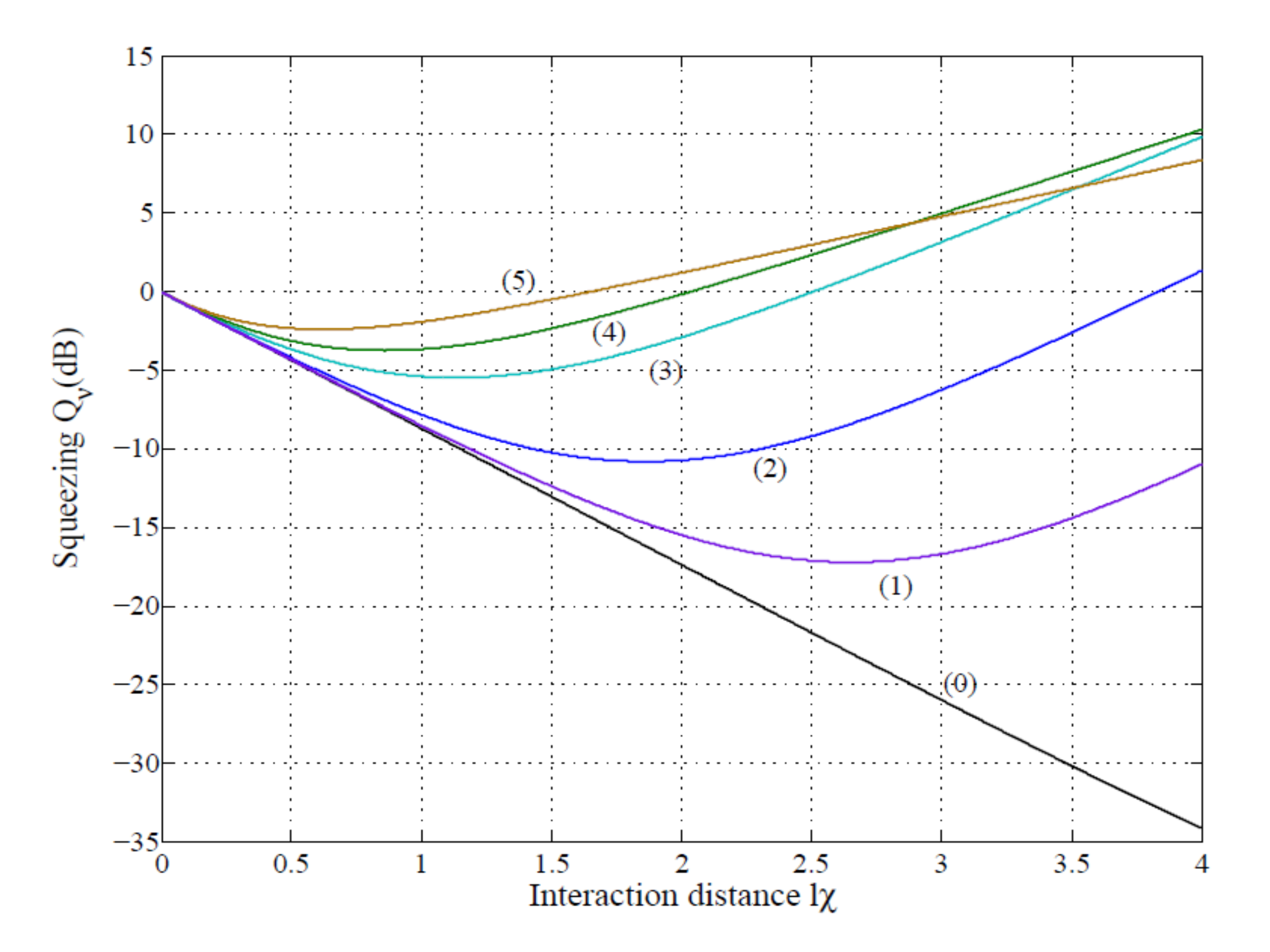}
\caption{\label{figureL} (Color online) Two-mode noise level \(Q_\nu\) (in dB) for the phase \(\Psi=\pi/2\) versus normalized length of interaction for zero detuning \(\nu=0\), zero temperature \(T=0\), and different parameters \(\kappa_\alpha/\chi\): (0) \(\kappa_\alpha/\chi=0\), (1) \(\kappa_\alpha/\chi=0.02\), (2) \(\kappa_\alpha/\chi=0.1\), (3) \(\kappa_\alpha/\chi=0.5\), (4) \(\kappa_\alpha/\chi=1\), (5) \(\kappa_\alpha/\chi=2\). }
\end{figure} 
\begin{figure*}
\includegraphics[width=\columnwidth, keepaspectratio=true]{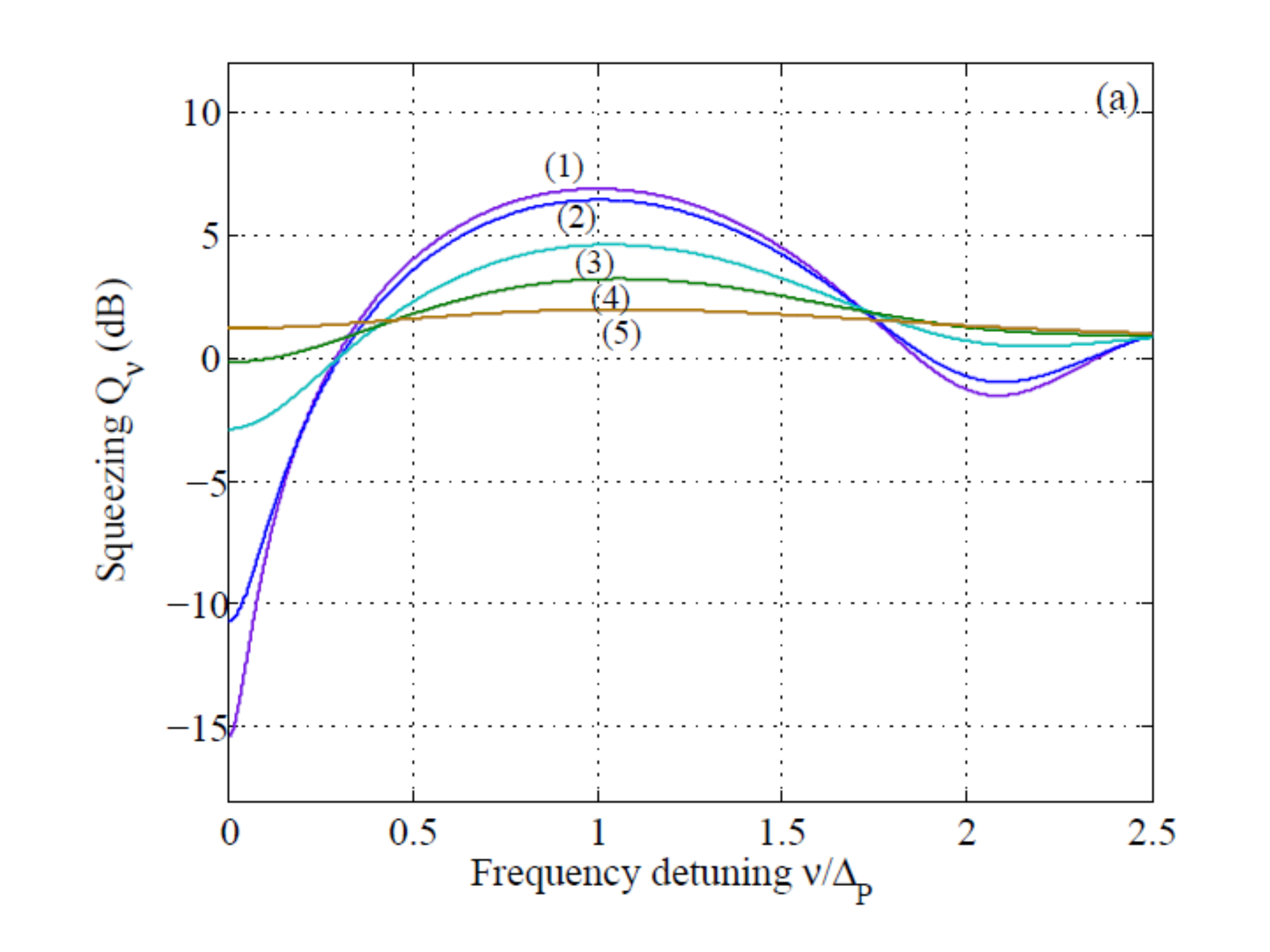}
\hfill
\includegraphics[width=\columnwidth, keepaspectratio=true]{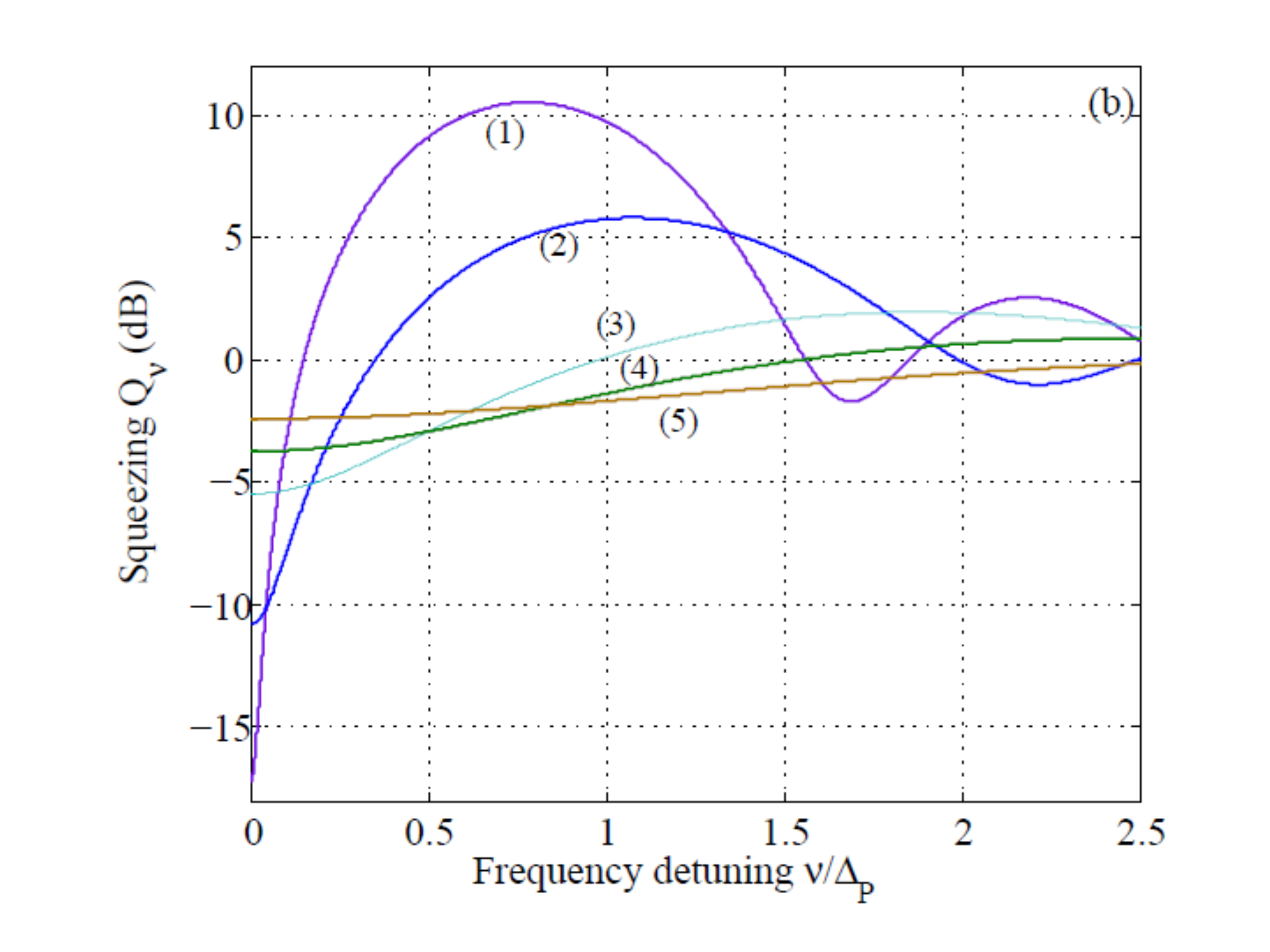}
\caption{\label{figureDis} (Color online) Two-mode noise level \(Q_\nu\) (in dB) for the phase \(\Psi=\pi/2\) versus normalized detuning, for zero temperature \(T=0\), and different parameters \(\kappa_\alpha(\nu=0)/\chi\): (1) \(\kappa_\alpha/\chi=0.02\), (2) \(\kappa_\alpha/\chi=0.1\), (3) \(\kappa_\alpha/\chi=0.5\), (4) \(\kappa_\alpha/\chi=1\), (5) \(\kappa_\alpha/\chi=2\). In Fig (a) all spectra are obtained at the same normalized length of interaction \(l\chi=2\). In Fig (b) each spectrum is obtained at the optimal length of interaction: (1) \(l\chi=2.65\), (2) \(l\chi=1.87\), (3) \(l\chi=1.1\), (4) \(l\chi=0.85\), (5) \(l\chi=0.62\). }
\end{figure*} 

How does the dissipation damage squeezing? The dissipation in the medium ``delivers'' an additional uncorrelated vacuum noise with amplitude, proportional to the absorption coefficient, and ``spoils'' correlations between two waves responsible for squeezing. Note, that the second effect does not appear for the symmetric equations for two waves, unlike the system under consideration.
Formally, nonzero partial decrement of \(\alpha\)-wave \(\kappa_\alpha\neq 0\) changes the ratio between partial amplitudes in normal waves \(\tilde{K}_0\) in such a way that the growing term in Eq.~\formref{full} can not be cut off by choosing appropriate phase \(\Psi\) any more. For the zero temperature (\(S_\alpha=0\)), at zero detuning \(\nu=0\) and \(\Psi=\pi/2\) we get from Eq.~\formref{full} the following relation:
\begin{eqnarray}
Q_{\nu=0}=\nonumber\\
\frac{1}{2}e^{-\kappa_\alpha l+2\chi\sqrt{1+\brac{\frac{\kappa_\alpha}{2\chi}}^2}l}\times\nonumber\\
\brac{1-\frac{1}{\sqrt{1+\brac{\frac{\kappa_\alpha}{2\chi}}^2}}}\brac{1+\frac{\kappa_\alpha}{2\chi\sqrt{1+\brac{\frac{\kappa_\alpha}{2\chi}}^2}}}+\nonumber\\
\frac{1}{2}e^{-\kappa_\alpha l-2\chi\sqrt{1+\brac{\frac{\kappa_\alpha}{2\chi}}^2}l}\times\nonumber\\
\brac{1+\frac{1}{\sqrt{1+\brac{\frac{\kappa_\alpha}{2\chi}}^2}}}\brac{1-\frac{\kappa_\alpha}{2\chi\sqrt{1+\brac{\frac{\kappa_\alpha}{2\chi}}^2}}}+\nonumber\\
e^{-\kappa_\alpha l}\frac{\kappa_\alpha}{2\chi\brac{1+\brac{\frac{\kappa_\alpha}{2\chi}}^2}}\nonumber\\
\label{noise_T_0}
\end{eqnarray} 
Even in the center of line the noise level decreases with the length of interaction only to a certain point after which it increases exponentially.
So there is an optimal length \(l_{opt}\) dependent on \(\kappa_\alpha\).  
In case of weak dissipation (or strong driving field) \(\kappa_\alpha\ll\chi\) (see Eq.~\formref{weak_diss}) we get from Eq.~\formref{noise_T_0}:
\begin{eqnarray}
l_{opt}\approx\frac{1}{2\chi}ln\frac{4\chi}{\kappa_\alpha}.
\label{l_opt_1}
\end{eqnarray}
The level of noise at this length is equal to
\begin{eqnarray}
Q_{\nu=0}(l_{opt})\approx\frac{\kappa_\alpha}{\chi}e^{-\kappa_\alpha l_{opt}}\approx\frac{\kappa_\alpha}{\chi}\brac{1+\frac{\kappa_\alpha}{2\chi}ln\frac{\kappa_\alpha}{4\chi}}\ll 1.\nonumber
\end{eqnarray}
The maximal length for which \(Q_{\nu=0}<1\) is \(l_{max}=2l_{opt}\).
The width of squeezing band for the optimal length can be estimated by Eq.~\formref{width}:
\begin{eqnarray}
\Delta_{sq}\approx {\Delta_P}\sqrt{\frac{\kappa_\alpha}{\chi}}=2\abs{\Omega}\sqrt{\frac{\gamma_{21}^0}.{\omega_{21}}}\nonumber
\end{eqnarray}

It should be noted that rather weak squeezing is realised for strong dissipation (or weak driving field), when the condition Eq.~\formref{weak_diss} is not fulfilled. So under condition \(\kappa_\alpha\gg\chi\) we get for the optimal length the following relation:
\begin{eqnarray}
l_{opt}\approx\frac{1}{\kappa_\alpha}ln\frac{\kappa_\alpha}{\chi}.
\label{l_opt_2}
\end{eqnarray}
The level of noise at this length is equal to:
\begin{eqnarray}
Q_{\nu=0}(l_{opt})\approx 1-\frac{2\chi}{\kappa_p}
\nonumber
\end{eqnarray}
and the frequency band of such squeezing is significantly higher than the frequency band of parametric instability.

Fig.~\ref{figureL} demonstrates the interaction length dependence of squeezing in the center of line. The noise spectra for different parameters \(\kappa_\alpha(\nu=0)/\chi\) and different lengths of interaction are shown in Fig.~\ref{figureDis}. The calculations hereinafter are made for parameter \(\frac{\abs{\Omega}}{\omega_{21}}\sqrt{\frac{\gamma_{31}}{\gamma_{21}^0}}=0.1\). On the one hand it provides the fulfilment of condition Eq.~\formref{small}, and on the other hand it should be taken into account to calculate correctly the influence of EIT resonance line shape within considered frequency domain.

It should be noted, that existence of the optimal length of interaction dependent on the partial absorption coefficient of \(\alpha\)-wave was indirectly confirmed experimentally in paper \cite{McCormick} where it was shown that for a given transmission of \(\alpha\)-wave there is an optimal parametric gain for the best squeezing based on 4WM in an atomic vapor. 

\subsection{Thermal fluctuations squeezing}

The presence of the thermal fluctuations in the medium may cardinally disturb generation of squeezed state of light. 

In accordance with general relation Eq.~\formref{full} the only parameter that changes the noise level of joint quadrature operator \(Q_{\nu}\)  under condition of thermal excitations 
in the medium is \(S_\alpha\) given by Eq.~\formref{S_p}, that in the considered nonlinear system is defined by the averaged number of thermal quanta at the low frequency of ground state splitting \(n_T(\omega_{21})\) and depends on rate of spontaneous emission and elastic dephasing rate at this transition.


Unlike the zero-temperature regime, when for every arbitrarily small parameter \({\chi}/{\kappa_\alpha}\) (in frame of mentioned restrictions) the squeezing takes place, in the presence of thermal fluctuations there is a threshold value for parameter \({\chi}/{\kappa_\alpha}>\brac{{\chi}/{\kappa_\alpha}}_{thr}\) dependent on \(S_\alpha\). The analysis of Eq.~\formref{full} for \(\nu=0\) has shown that 
the noise level at the optimal value for density-length product, that depends on the parameter \({\chi}/{\kappa_\alpha}\) (in the extreme cases given by relations Eqs.~\formref{l_opt_1},\formref{l_opt_2}) 
is modified in a simple way:
\begin{eqnarray}
Q_{\nu=0}(l_{opt})\approx Q_{\nu=0}(l_{opt})\vert_{S_\alpha=0}\times\brac{S_\alpha+1}.\nonumber
\end{eqnarray}
Using this relation it can be shown that in case of low temperature, when \(S_\alpha\ll 1\), the threshold value is low, defined by the following expression:  
\begin{eqnarray}
\brac{\frac{\chi}{\kappa_\alpha}}_{thr}\approx S_\alpha/2.\nonumber
\end{eqnarray}
And for strong parametric coupling regime when \({\chi}/{\kappa_\alpha}\gg 1\) the presence of thermal fluctuations weekly changes both peak value of squeezing and the noise spectrum.

In case \(S_\alpha\gtrsim 1\) the threshold value becomes essential:
\begin{eqnarray}
\brac{\frac{\chi}{\kappa_\alpha}}_{thr}\approx S_\alpha \nonumber
\end{eqnarray}
and even in the optimal conditions (\(\chi\gg\kappa_\alpha\)) the noise level increases in \(S_\alpha+1\) times in comparison with zero-temperature regime:
\begin{eqnarray}
Q_{\nu=0}(l_{opt})\approx \frac{\kappa_\alpha}{\chi}\times\brac{S_\alpha+1}.
\label{final}
\end{eqnarray}

\begin{figure*}
\includegraphics[width=0.68\columnwidth, keepaspectratio=true]{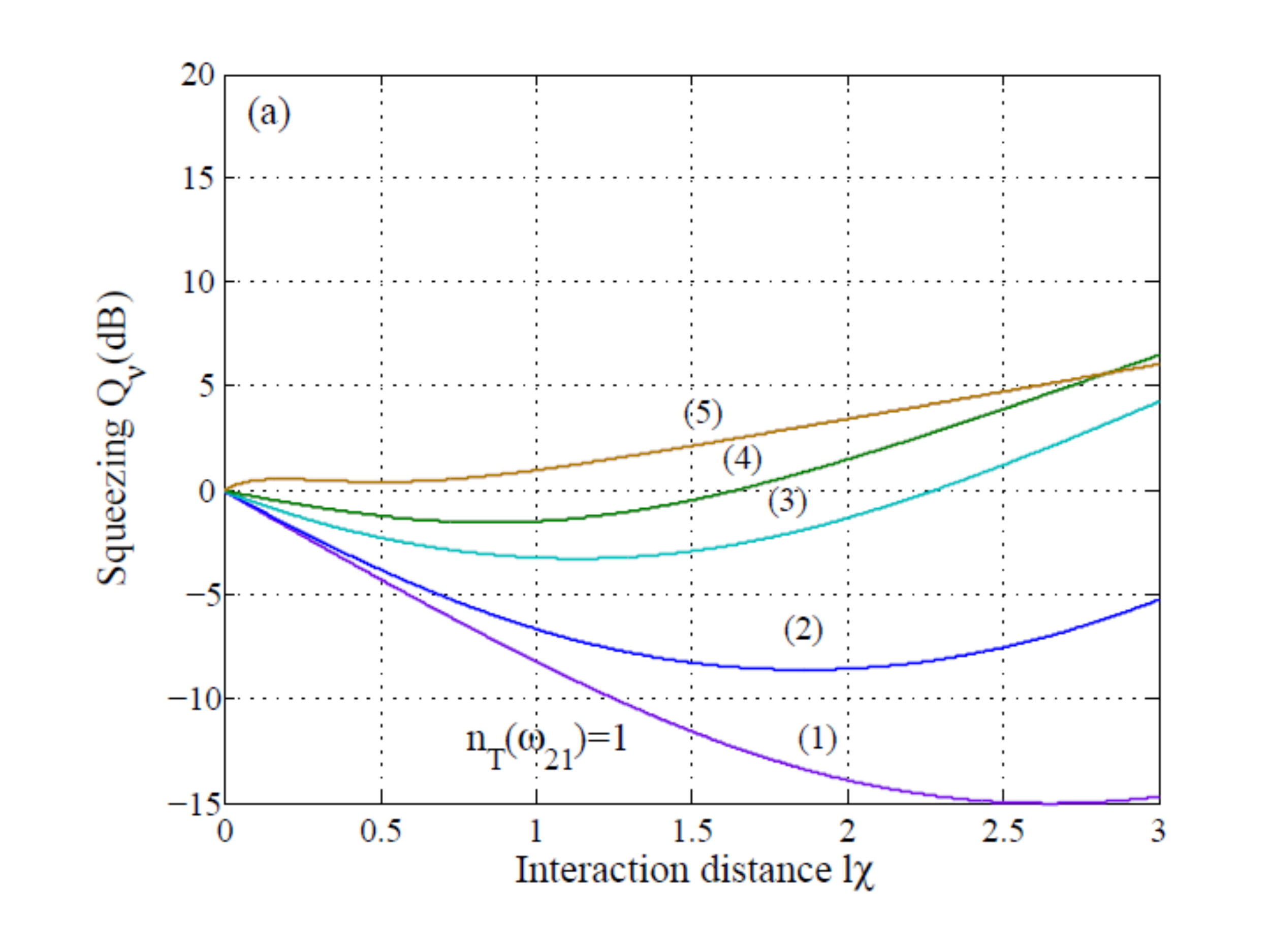}
\hfill
\includegraphics[width=0.68\columnwidth, keepaspectratio=true]{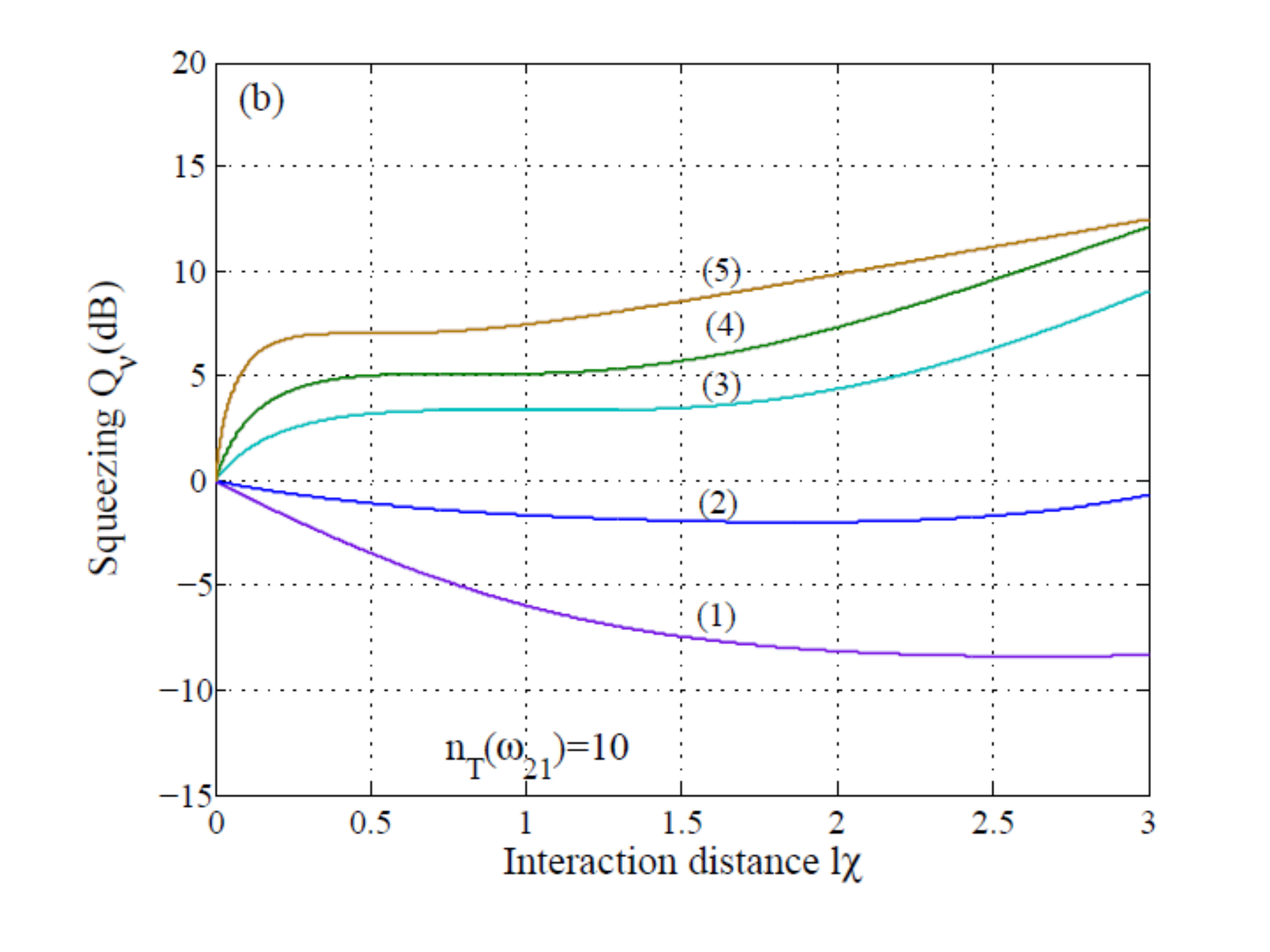}
\hfill
\includegraphics[width=0.68\columnwidth, keepaspectratio=true]{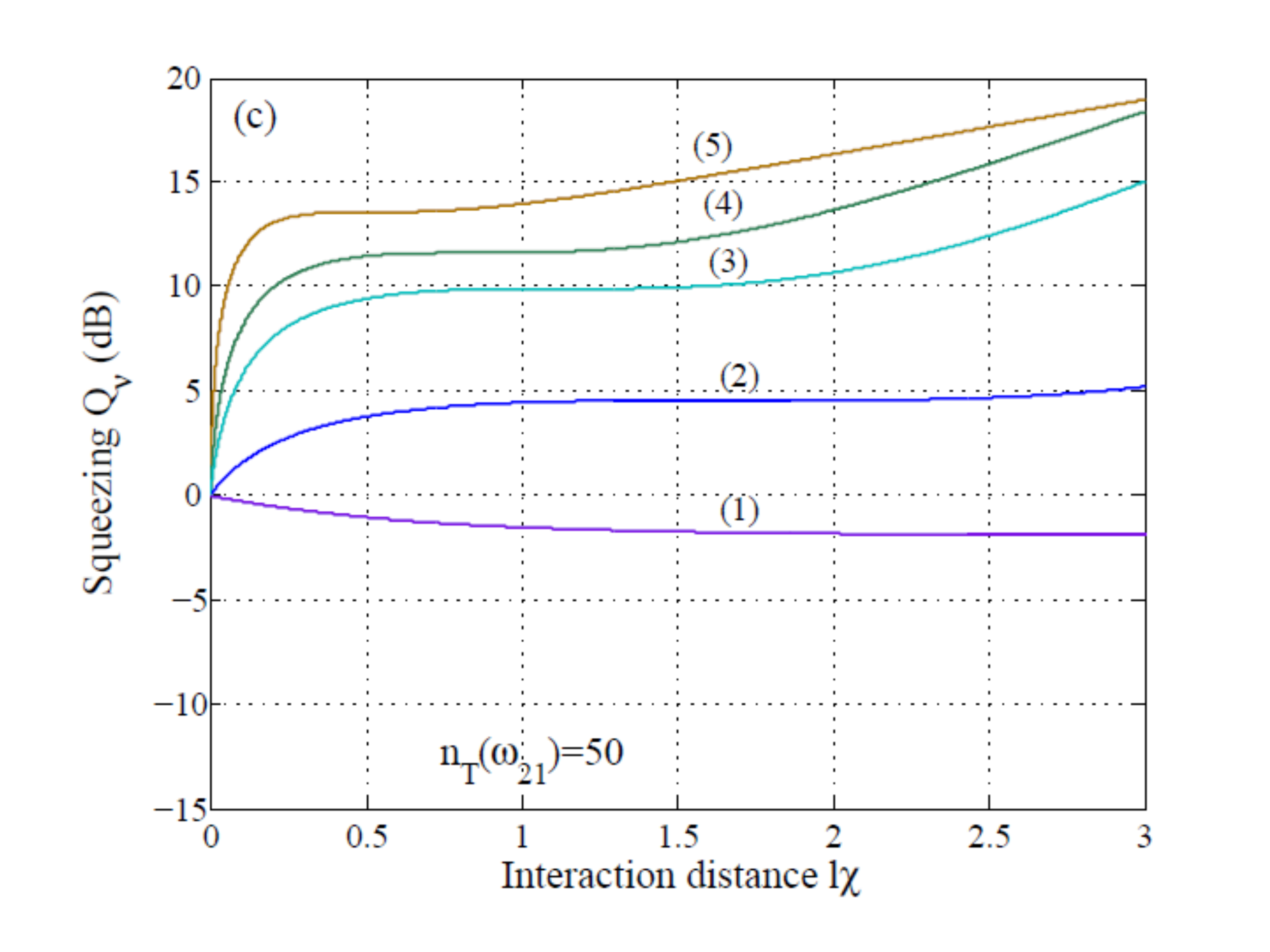}
\\
\includegraphics[width=0.68\columnwidth, keepaspectratio=true]{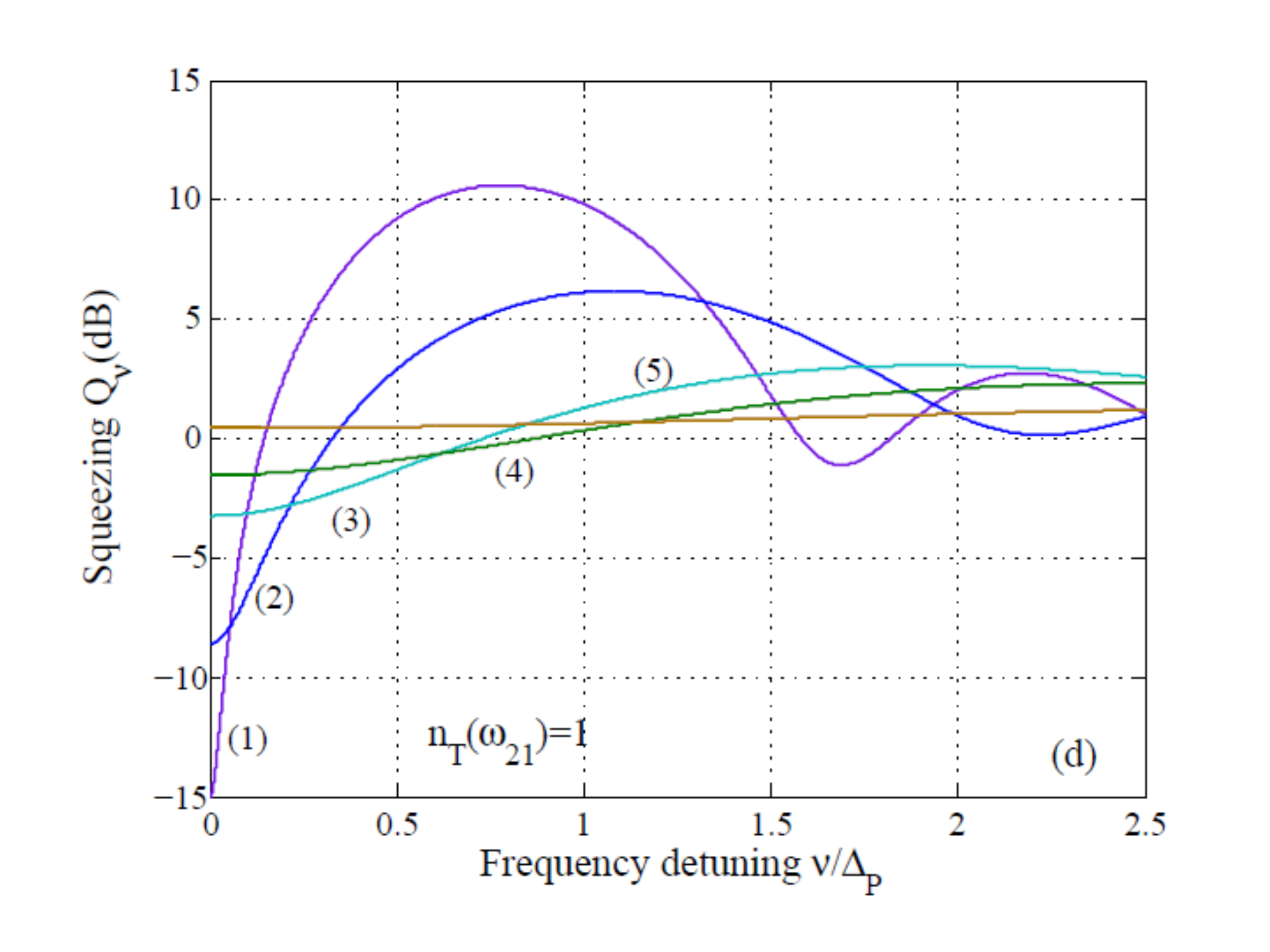}
\hfill
\includegraphics[width=0.68\columnwidth, keepaspectratio=true]{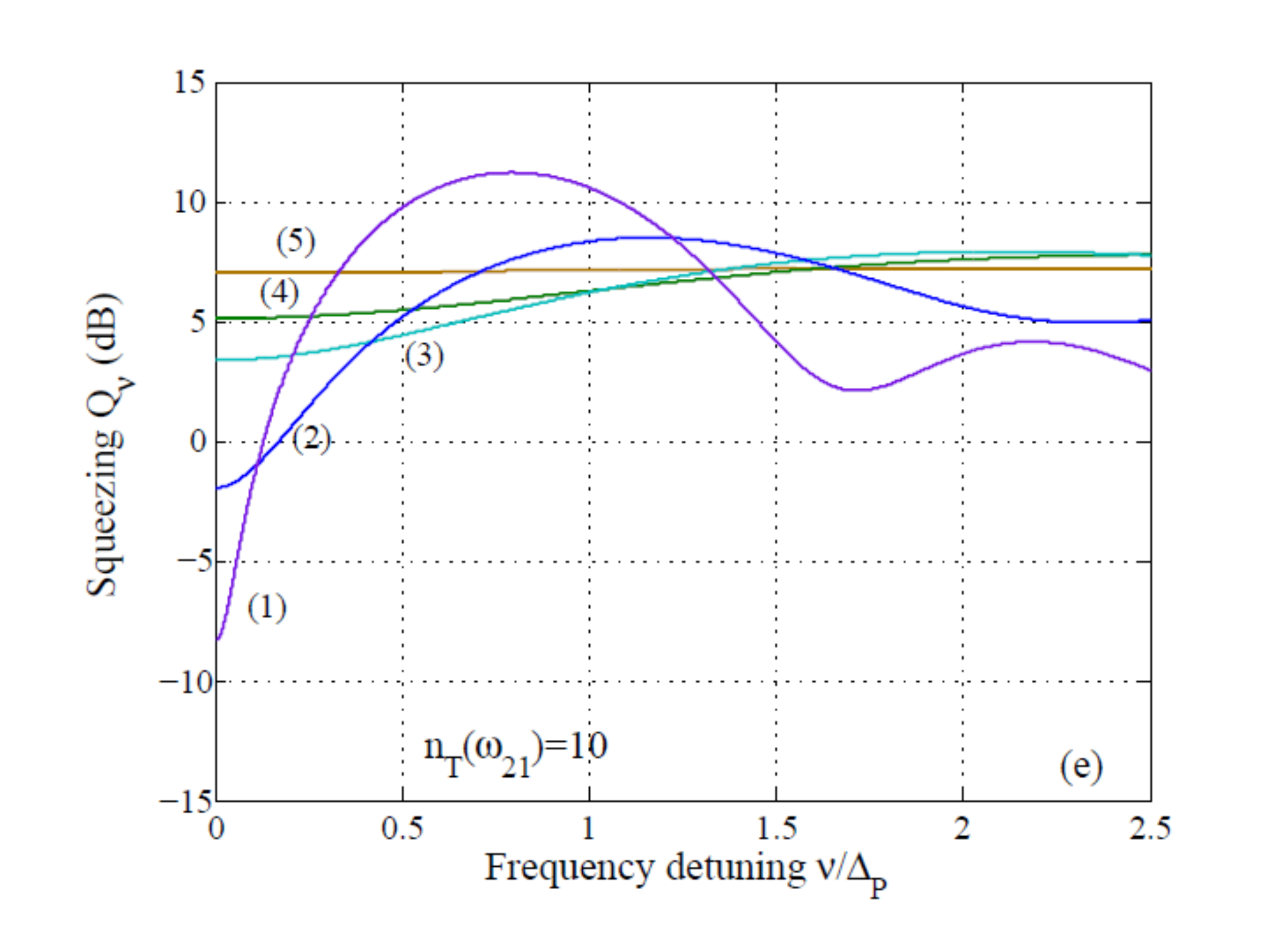}
\hfill
\includegraphics[width=0.68\columnwidth, keepaspectratio=true]{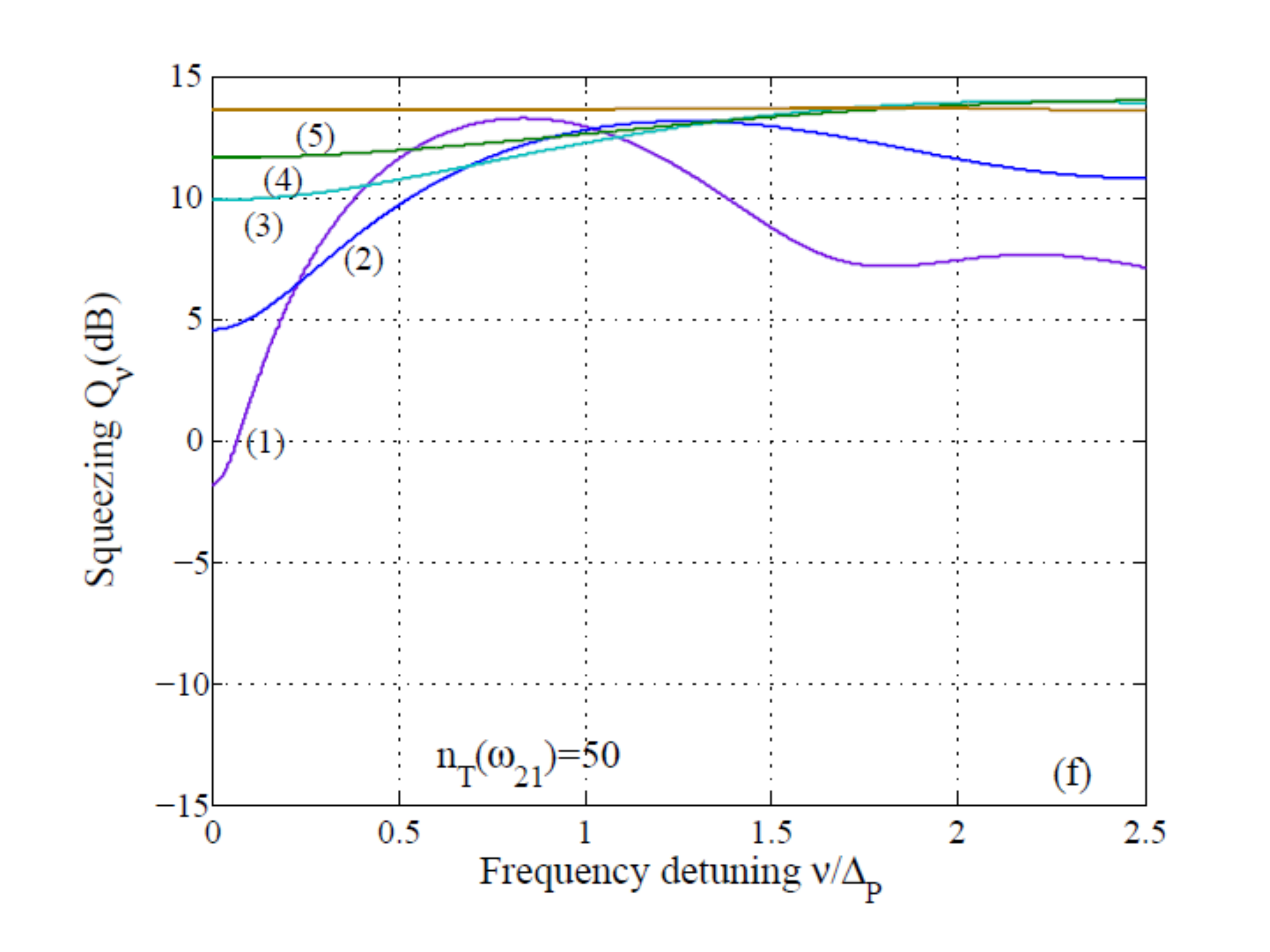}
\caption{\label{figureT} (Color online) Two-mode noise level \(Q_\nu\) (in dB) for the phase \(\Psi=\pi/2\) for different temperatures (the corresponding averaged number of thermal photos at frequency \(\omega_{21}\) is pointed at every figure), and different parameters \(\kappa_\alpha(\nu=0)/\chi\): (1) \(\kappa_\alpha/\chi=0.02\), (2) \(\kappa_\alpha/\chi=0.1\), (3) \(\kappa_\alpha/\chi=0.5\), (4) \(\kappa_\alpha/\chi=1\), (5) \(\kappa_\alpha/\chi=3\). In Figures (a-c) the length dependence of the squeezing in the center of line is presented. In Figures (d-f) the noise spectra are presented, each spectrum is obtained at the optimal length of interaction: (1) \(l\chi=2.65\), (2) \(l\chi=1.87\), (3) \(l\chi=1.1\), (4) \(l\chi=0.85\), (5) \(l\chi=0.5\).}
\end{figure*}

Fig.~\ref{figureT} presents the noise spectrum and dependence of squeezing in the center of line on the density-length product for different values of 
\(n_T(\omega_{21})\).  
The ratio of spontaneous emission rate at the transition \(\cet{1}-\cet{2}\) to the rate of coherence relaxation at this transition caused by elastic processes \(A_{21}/\Gamma_{21}\) is taken equal to \(1\) so that \(S_{\alpha}(\nu=0)=\frac{2}{3}n_T(\omega_{21})\). 

\section{Discussion}
We presented here the detailed analytical investigation of two-mode squeezed vacuum generation in robust scheme of four-waves mixing in a resonant \(\Lambda\)-configuration taking into account such negative factors, as dissipation caused by relaxation in atomic system and thermal excitations delivering the additional uncorrelated noise. These processes are considered on the basis of self-consistent microscopic approach. We 
investigated the influence of spontaneous Raman scattering under resonant conditions of EIT on the level of two-mode squeezing. We obtained the analytical formulas for the optimal density-length product of atomic medium and for the frequency width of squeezing band as they depend on the drive intensity and the relaxation rate.

The following illustrative estimations of real experimental parameters, for example for Rb vapour (wave length of resonant radiation \(\lambda =794 \mathrm{nm}\), the frequency of ground state splitting \(\omega_{21}=8.83\mathrm{GHz}\)), can be made. For the relaxation rates \(\gamma_{31}=100\mathrm{MHz}\), \(\gamma_{21}=15\mathrm{kHz}\),
if the drive power is about \(10\mathrm{mW}\) and the beam focusing diameter is \(2\mathrm{mm}\) (so that Rabi frequency \(\abs{\Omega}=28\mathrm{MHz}\))  the condition of strong parametric instability Eq.~\formref{weak_diss} is fulfilled: \(\chi/\kappa_\alpha=8\). Then calculated dimensionless optimal density-length product \(l_{opt}\chi=1.7\) corresponds to the following dimension value: \(lN=5.2\mathrm{cm}\times 6.6\times 10^{11}\mathrm{cm}^{-3}\). The level of squeezing in the center of line in ``cold'' conditions is equal to \(-10\mathrm{dB}\) and the frequency band of squeezing \(\Delta_{sq}=83 \mathrm{kHz}\). How can this result be spoiled in  ``hot'' conditions? It can be easily shown that the number of thermal photons at frequency \(\omega_{21}\) is equal to \(1\) for temperature \(T=0.5\mathrm{K}\), while at room temperature \(T=290\mathrm{K}\) it is already about \(1000\). But the parameter \(S_{\alpha}\) is not as much as \(n_T(\omega_{21})\) since the rate of coherence decay at the transition \(\cet{2}-\cet{1}\) in a greater degree is determined by elastic processes. But if the rate of spontaneous emission \(A_{21}\) is at least 
\(0.01\Gamma_{21}\) then \(S_\alpha>10\), that in \(10\) times increases noise of light at optical transition with respect to zero-temperature regime.

It should be noted that in the above discussion of temperature dependence of squeezing level we did not take into account that the dephasing rate \(\Gamma_{21}\) increases with \(T\), that will formally lead to effective reduction of parameter \(S_\alpha\), denoting the share of  ``Raman noise'' in the total noise level. But this total noise level increases with a rise in a dephasing rate at low-frequency transition. 
It follows from Eq.~\formref{final}, that taking into account the temperature dependence of \(\Gamma_{21}\) we get additional negative factor:
\begin{eqnarray}
\frac{Q(T)}{Q(T=0)}=1+\frac{\Delta\Gamma_{21}}{\Gamma_{21}^0+\frac{1}{2}A_{21}}+S_\alpha(\Gamma_{21}^0),\nonumber
\end{eqnarray} 
where \(\Delta\Gamma_{21}=\Gamma_{21}(T)-\Gamma_{21}^0\), \(\Gamma_{21}^0=\Gamma_{21}(T=0)\). Also we get that the damage factor associated with the thermal excitations in the medium depends on the ratio of spontaneous emission rate \(A_{21}\) to the dephasing rate \(\Gamma_{21}\) at zero temperature, this ratio can be significant.

So the following general conclusion can be made. 
Strong parametric coupling in regime of four-wave mixing in a \(\Lambda\)-scheme of three level atoms may provide two-mode squeezing not only of the intrinsic quantum fluctuations of light but also squeezing of thermal fluctuations. But conditions of effective squeezing is much more demanding if number of thermal photons at low frequency of ground state splitting is high and a role of spontaneous emission process in ground state coherence decay is not negligible.

\begin{acknowledgments}
This work was supported by RFBR grants N. 14-29-07152, 14-22-02034.
\end{acknowledgments}

\bibliography{Erukhimova}

\providecommand{\noopsort}[1]{}\providecommand{\singleletter}[1]{#1}%
\begin{thebibliography}{38}%
\makeatletter
\providecommand \@ifxundefined [1]{%
 \@ifx{#1\undefined}
}%
\providecommand \@ifnum [1]{%
 \ifnum #1\expandafter \@firstoftwo
 \else \expandafter \@secondoftwo
 \fi
}%
\providecommand \@ifx [1]{%
 \ifx #1\expandafter \@firstoftwo
 \else \expandafter \@secondoftwo
 \fi
}%
\providecommand \natexlab [1]{#1}%
\providecommand \enquote  [1]{``#1''}%
\providecommand \bibnamefont  [1]{#1}%
\providecommand \bibfnamefont [1]{#1}%
\providecommand \citenamefont [1]{#1}%
\providecommand \href@noop [0]{\@secondoftwo}%
\providecommand \href [0]{\begingroup \@sanitize@url \@href}%
\providecommand \@href[1]{\@@startlink{#1}\@@href}%
\providecommand \@@href[1]{\endgroup#1\@@endlink}%
\providecommand \@sanitize@url [0]{\catcode `\\12\catcode `\$12\catcode
  `\&12\catcode `\#12\catcode `\^12\catcode `\_12\catcode `\%12\relax}%
\providecommand \@@startlink[1]{}%
\providecommand \@@endlink[0]{}%
\providecommand \url  [0]{\begingroup\@sanitize@url \@url }%
\providecommand \@url [1]{\endgroup\@href {#1}{\urlprefix }}%
\providecommand \urlprefix  [0]{URL }%
\providecommand \Eprint [0]{\href }%
\providecommand \doibase [0]{http://dx.doi.org/}%
\providecommand \selectlanguage [0]{\@gobble}%
\providecommand \bibinfo  [0]{\@secondoftwo}%
\providecommand \bibfield  [0]{\@secondoftwo}%
\providecommand \translation [1]{[#1]}%
\providecommand \BibitemOpen [0]{}%
\providecommand \bibitemStop [0]{}%
\providecommand \bibitemNoStop [0]{.\EOS\space}%
\providecommand \EOS [0]{\spacefactor3000\relax}%
\providecommand \BibitemShut  [1]{\csname bibitem#1\endcsname}%
\let\auto@bib@innerbib\@empty
\bibitem [{\citenamefont {Slusher}\ \emph {et~al.}(1985)\citenamefont {Slusher}
  \emph {et~al.}}]{Slusher}%
  \BibitemOpen
  \bibfield  {author} {\bibinfo {author} {\bibfnamefont {R.~E.}\ \bibnamefont
  {Slusher}} \emph {et~al.},\ }\href@noop {} {\bibfield  {journal} {\bibinfo
  {journal} {Phys. Rev. Lett}\ }\textbf {\bibinfo {volume} {55}},\ \bibinfo
  {pages} {2409} (\bibinfo {year} {1985})}\BibitemShut {NoStop}%
\bibitem [{\citenamefont {Maeda}\ \emph {et~al.}(1987)\citenamefont {Maeda},
  \citenamefont {Kumar},\ and\ \citenamefont {Shapiro}}]{Maeda}%
  \BibitemOpen
  \bibfield  {author} {\bibinfo {author} {\bibfnamefont {M.~W.}\ \bibnamefont
  {Maeda}}, \bibinfo {author} {\bibfnamefont {P.}~\bibnamefont {Kumar}}, \ and\
  \bibinfo {author} {\bibfnamefont {I.~H.}\ \bibnamefont {Shapiro}},\
  }\href@noop {} {\bibfield  {journal} {\bibinfo  {journal} {Opt. Lett.}\
  }\textbf {\bibinfo {volume} {12}},\ \bibinfo {pages} {161} (\bibinfo {year}
  {1987})}\BibitemShut {NoStop}%
\bibitem [{\citenamefont {Vallet}\ \emph {et~al.}(1990)\citenamefont {Vallet},
  \citenamefont {Pinard},\ and\ \citenamefont {Grynberg}}]{Vallet}%
  \BibitemOpen
  \bibfield  {author} {\bibinfo {author} {\bibfnamefont {M.}~\bibnamefont
  {Vallet}}, \bibinfo {author} {\bibfnamefont {M.}~\bibnamefont {Pinard}}, \
  and\ \bibinfo {author} {\bibfnamefont {G.}~\bibnamefont {Grynberg}},\
  }\href@noop {} {\bibfield  {journal} {\bibinfo  {journal} {Europhys. Lett.}\
  }\textbf {\bibinfo {volume} {11}},\ \bibinfo {pages} {739} (\bibinfo {year}
  {1990})}\BibitemShut {NoStop}%
\bibitem [{\citenamefont {Lambrecht}\ \emph {et~al.}(1996)\citenamefont
  {Lambrecht} \emph {et~al.}}]{Lambrecht}%
  \BibitemOpen
  \bibfield  {author} {\bibinfo {author} {\bibfnamefont {A.}~\bibnamefont
  {Lambrecht}} \emph {et~al.},\ }\href@noop {} {\bibfield  {journal} {\bibinfo
  {journal} {Europhys. Lett.}\ }\textbf {\bibinfo {volume} {36}},\ \bibinfo
  {pages} {93} (\bibinfo {year} {1996})}\BibitemShut {NoStop}%
\bibitem [{\citenamefont {Josse}\ \emph {et~al.}(2003)\citenamefont {Josse}
  \emph {et~al.}}]{Josse}%
  \BibitemOpen
  \bibfield  {author} {\bibinfo {author} {\bibfnamefont {V.}~\bibnamefont
  {Josse}} \emph {et~al.},\ }\href@noop {} {\bibfield  {journal} {\bibinfo
  {journal} {Phys. Rev. Lett.}\ }\textbf {\bibinfo {volume} {91}},\ \bibinfo
  {pages} {103601} (\bibinfo {year} {2003})}\BibitemShut {NoStop}%
\bibitem [{\citenamefont {Grove}\ \emph {et~al.}(1997)\citenamefont {Grove}
  \emph {et~al.}}]{Grove}%
  \BibitemOpen
  \bibfield  {author} {\bibinfo {author} {\bibfnamefont {T.~T.}\ \bibnamefont
  {Grove}} \emph {et~al.},\ }\href@noop {} {\bibfield  {journal} {\bibinfo
  {journal} {Opt. Lett.}\ }\textbf {\bibinfo {volume} {22}},\ \bibinfo {pages}
  {769} (\bibinfo {year} {1997})}\BibitemShut {NoStop}%
\bibitem [{\citenamefont {Shahriar}\ and\ \citenamefont
  {Hemmer}(1998)}]{Shahriar}%
  \BibitemOpen
  \bibfield  {author} {\bibinfo {author} {\bibfnamefont {M.}~\bibnamefont
  {Shahriar}}\ and\ \bibinfo {author} {\bibfnamefont {P.}~\bibnamefont
  {Hemmer}},\ }\href@noop {} {\bibfield  {journal} {\bibinfo  {journal} {Opt.
  Comm.}\ }\textbf {\bibinfo {volume} {158}},\ \bibinfo {pages} {273} (\bibinfo
  {year} {1998})}\BibitemShut {NoStop}%
\bibitem [{\citenamefont {Lukin}\ \emph {et~al.}(1999)\citenamefont {Lukin}
  \emph {et~al.}}]{Lukin_Matsko}%
  \BibitemOpen
  \bibfield  {author} {\bibinfo {author} {\bibfnamefont {M.~D.}\ \bibnamefont
  {Lukin}} \emph {et~al.},\ }\href@noop {} {\bibfield  {journal} {\bibinfo
  {journal} {Phys. Rev. Lett.}\ }\textbf {\bibinfo {volume} {82}},\ \bibinfo
  {pages} {1847} (\bibinfo {year} {1999})}\BibitemShut {NoStop}%
\bibitem [{\citenamefont {Balic}\ \emph {et~al.}(2005)\citenamefont {Balic},
  \citenamefont {Braje}, \citenamefont {Kolchin}, \citenamefont {Yin},\ and\
  \citenamefont {Harris}}]{Balic}%
  \BibitemOpen
  \bibfield  {author} {\bibinfo {author} {\bibfnamefont {V.}~\bibnamefont
  {Balic}}, \bibinfo {author} {\bibfnamefont {D.~A.}\ \bibnamefont {Braje}},
  \bibinfo {author} {\bibfnamefont {P.}~\bibnamefont {Kolchin}}, \bibinfo
  {author} {\bibfnamefont {G.~Y.}\ \bibnamefont {Yin}}, \ and\ \bibinfo
  {author} {\bibfnamefont {S.~E.}\ \bibnamefont {Harris}},\ }\href@noop {}
  {\bibfield  {journal} {\bibinfo  {journal} {Phys. Rev. Lett.}\ }\textbf
  {\bibinfo {volume} {94}},\ \bibinfo {pages} {183601} (\bibinfo {year}
  {2005})}\BibitemShut {NoStop}%
\bibitem [{\citenamefont {Kolchin}\ \emph {et~al.}(2006)\citenamefont {Kolchin}
  \emph {et~al.}}]{Kolchin06}%
  \BibitemOpen
  \bibfield  {author} {\bibinfo {author} {\bibfnamefont {P.}~\bibnamefont
  {Kolchin}} \emph {et~al.},\ }\href@noop {} {\bibfield  {journal} {\bibinfo
  {journal} {Phys. Rev. Lett.}\ }\textbf {\bibinfo {volume} {97}},\ \bibinfo
  {pages} {113602} (\bibinfo {year} {2006})}\BibitemShut {NoStop}%
\bibitem [{\citenamefont {Kolchin}(2007)}]{Kolchin07}%
  \BibitemOpen
  \bibfield  {author} {\bibinfo {author} {\bibfnamefont {P.}~\bibnamefont
  {Kolchin}},\ }\href@noop {} {\bibfield  {journal} {\bibinfo  {journal} {Phys.
  Rev. A}\ }\textbf {\bibinfo {volume} {75}},\ \bibinfo {pages} {033814}
  (\bibinfo {year} {2007})}\BibitemShut {NoStop}%
\bibitem [{\citenamefont {McCormick}\ \emph {et~al.}(2007)\citenamefont
  {McCormick} \emph {et~al.}}]{McCormickOL}%
  \BibitemOpen
  \bibfield  {author} {\bibinfo {author} {\bibfnamefont {C.~F.}\ \bibnamefont
  {McCormick}} \emph {et~al.},\ }\href@noop {} {\bibfield  {journal} {\bibinfo
  {journal} {Optics Letters}\ }\textbf {\bibinfo {volume} {32}},\ \bibinfo
  {pages} {178} (\bibinfo {year} {2007})}\BibitemShut {NoStop}%
\bibitem [{\citenamefont {McCormick}\ \emph {et~al.}(2008)\citenamefont
  {McCormick} \emph {et~al.}}]{McCormick}%
  \BibitemOpen
  \bibfield  {author} {\bibinfo {author} {\bibfnamefont {C.~F.}\ \bibnamefont
  {McCormick}} \emph {et~al.},\ }\href@noop {} {\bibfield  {journal} {\bibinfo
  {journal} {Phys. Rev. A}\ }\textbf {\bibinfo {volume} {78}},\ \bibinfo
  {pages} {043816} (\bibinfo {year} {2008})}\BibitemShut {NoStop}%
\bibitem [{\citenamefont {Boyer}\ \emph {et~al.}(2008)\citenamefont {Boyer}
  \emph {et~al.}}]{BoyerScience}%
  \BibitemOpen
  \bibfield  {author} {\bibinfo {author} {\bibfnamefont {V.}~\bibnamefont
  {Boyer}} \emph {et~al.},\ }\href@noop {} {\bibfield  {journal} {\bibinfo
  {journal} {Science}\ }\textbf {\bibinfo {volume} {321}},\ \bibinfo {pages}
  {544} (\bibinfo {year} {2008})}\BibitemShut {NoStop}%
\bibitem [{\citenamefont {Glorieux}\ \emph {et~al.}(2010)\citenamefont
  {Glorieux} \emph {et~al.}}]{Glorieux}%
  \BibitemOpen
  \bibfield  {author} {\bibinfo {author} {\bibfnamefont {Q.}~\bibnamefont
  {Glorieux}} \emph {et~al.},\ }\href@noop {} {\bibfield  {journal} {\bibinfo
  {journal} {Phys. Rev. A}\ }\textbf {\bibinfo {volume} {82}},\ \bibinfo
  {pages} {033819} (\bibinfo {year} {2010})}\BibitemShut {NoStop}%
\bibitem [{\citenamefont {Qin}\ \emph {et~al.}(2014)\citenamefont {Qin} \emph
  {et~al.}}]{Qin}%
  \BibitemOpen
  \bibfield  {author} {\bibinfo {author} {\bibfnamefont {Z.}~\bibnamefont
  {Qin}} \emph {et~al.},\ }\href@noop {} {\bibfield  {journal} {\bibinfo
  {journal} {Phys. Rev. Lett.}\ }\textbf {\bibinfo {volume} {113}},\ \bibinfo
  {pages} {023602} (\bibinfo {year} {2014})}\BibitemShut {NoStop}%
\bibitem [{\citenamefont {Lvovsky}(2015)}]{Lvovsky}%
  \BibitemOpen
  \bibfield  {author} {\bibinfo {author} {\bibfnamefont {A.~I.}\ \bibnamefont
  {Lvovsky}},\ }\href@noop {} {\emph {\bibinfo {title} {Squeezed Light, in
  Photonics: Scientific Foundations, Technology and Applications, Volume 1 (ed
  D. L. Andrews)}}}\ (\bibinfo  {publisher} {John Wiley and Sons, Inc.},\
  \bibinfo {address} {Hoboken, NJ, USA},\ \bibinfo {year} {2015})\BibitemShut
  {NoStop}%
\bibitem [{\citenamefont {Callen}\ and\ \citenamefont {Welton}(1951)}]{Callen}%
  \BibitemOpen
  \bibfield  {author} {\bibinfo {author} {\bibfnamefont {H.~B.}\ \bibnamefont
  {Callen}}\ and\ \bibinfo {author} {\bibfnamefont {T.~A.}\ \bibnamefont
  {Welton}},\ }\href@noop {} {\bibfield  {journal} {\bibinfo  {journal} {Phys.
  Rev.}\ }\textbf {\bibinfo {volume} {83}},\ \bibinfo {pages} {34} (\bibinfo
  {year} {1951})}\BibitemShut {NoStop}%
\bibitem [{\citenamefont {Erukhimova}\ and\ \citenamefont
  {Tokman}(2015)}]{ErukhOL15}%
  \BibitemOpen
  \bibfield  {author} {\bibinfo {author} {\bibfnamefont {M.}~\bibnamefont
  {Erukhimova}}\ and\ \bibinfo {author} {\bibfnamefont {M.}~\bibnamefont
  {Tokman}},\ }\href@noop {} {\bibfield  {journal} {\bibinfo  {journal} {Optics
  Letters}\ }\textbf {\bibinfo {volume} {40}},\ \bibinfo {pages} {2739}
  (\bibinfo {year} {2015})}\BibitemShut {NoStop}%
\bibitem [{\citenamefont {Hsu}\ \emph {et~al.}(2006)\citenamefont {Hsu} \emph
  {et~al.}}]{Hsu}%
  \BibitemOpen
  \bibfield  {author} {\bibinfo {author} {\bibfnamefont {M.~T.~L.}\
  \bibnamefont {Hsu}} \emph {et~al.},\ }\href@noop {} {\bibfield  {journal}
  {\bibinfo  {journal} {Phys. Rev. Lett.}\ }\textbf {\bibinfo {volume} {97}},\
  \bibinfo {pages} {183601} (\bibinfo {year} {2006})}\BibitemShut {NoStop}%
\bibitem [{\citenamefont {Figueroa}\ \emph {et~al.}(2009)\citenamefont
  {Figueroa} \emph {et~al.}}]{Figueroa}%
  \BibitemOpen
  \bibfield  {author} {\bibinfo {author} {\bibfnamefont {E.}~\bibnamefont
  {Figueroa}} \emph {et~al.},\ }\href@noop {} {\bibfield  {journal} {\bibinfo
  {journal} {New Journal of Physics}\ }\textbf {\bibinfo {volume} {11}},\
  \bibinfo {pages} {013044} (\bibinfo {year} {2009})}\BibitemShut {NoStop}%
\bibitem [{\citenamefont {Hetet}\ \emph {et~al.}(2008)\citenamefont {Hetet}
  \emph {et~al.}}]{Hetet}%
  \BibitemOpen
  \bibfield  {author} {\bibinfo {author} {\bibfnamefont {G.}~\bibnamefont
  {Hetet}} \emph {et~al.},\ }\href@noop {} {\bibfield  {journal} {\bibinfo
  {journal} {Phys. Rev. A}\ }\textbf {\bibinfo {volume} {77}},\ \bibinfo
  {pages} {012323} (\bibinfo {year} {2008})}\BibitemShut {NoStop}%
\bibitem [{\citenamefont {Reim}\ \emph {et~al.}(2011)\citenamefont {Reim} \emph
  {et~al.}}]{Reim}%
  \BibitemOpen
  \bibfield  {author} {\bibinfo {author} {\bibfnamefont {K.~F.}\ \bibnamefont
  {Reim}} \emph {et~al.},\ }\href@noop {} {\bibfield  {journal} {\bibinfo
  {journal} {Phys. Rev. Lett.}\ }\textbf {\bibinfo {volume} {107}},\ \bibinfo
  {pages} {053603} (\bibinfo {year} {2011})}\BibitemShut {NoStop}%
\bibitem [{\citenamefont {Vdovin}\ and\ \citenamefont {Tokman}(2013)}]{Vdovin}%
  \BibitemOpen
  \bibfield  {author} {\bibinfo {author} {\bibfnamefont {V.}~\bibnamefont
  {Vdovin}}\ and\ \bibinfo {author} {\bibfnamefont {M.}~\bibnamefont
  {Tokman}},\ }\href@noop {} {\bibfield  {journal} {\bibinfo  {journal} {Phys.
  Rev. A}\ }\textbf {\bibinfo {volume} {87}},\ \bibinfo {pages} {012323}
  (\bibinfo {year} {2013})}\BibitemShut {NoStop}%
\bibitem [{\citenamefont {Tokman}\ \emph {et~al.}(2008)\citenamefont {Tokman},
  \citenamefont {Erukhimova},\ and\ \citenamefont {D'yachenko}}]{Tokman}%
  \BibitemOpen
  \bibfield  {author} {\bibinfo {author} {\bibfnamefont {M.~D.}\ \bibnamefont
  {Tokman}}, \bibinfo {author} {\bibfnamefont {M.~A.}\ \bibnamefont
  {Erukhimova}}, \ and\ \bibinfo {author} {\bibfnamefont {D.~O.}\ \bibnamefont
  {D'yachenko}},\ }\href@noop {} {\bibfield  {journal} {\bibinfo  {journal}
  {Phys. Rev. A.}\ }\textbf {\bibinfo {volume} {78}},\ \bibinfo {pages}
  {053808} (\bibinfo {year} {2008})}\BibitemShut {NoStop}%
\bibitem [{\citenamefont {Erukhimova}\ and\ \citenamefont
  {Tokman}(2011)}]{ErukhPRA11}%
  \BibitemOpen
  \bibfield  {author} {\bibinfo {author} {\bibfnamefont {M.}~\bibnamefont
  {Erukhimova}}\ and\ \bibinfo {author} {\bibfnamefont {M.}~\bibnamefont
  {Tokman}},\ }\href@noop {} {\bibfield  {journal} {\bibinfo  {journal} {Phys.
  Rev. A}\ }\textbf {\bibinfo {volume} {83}},\ \bibinfo {pages} {063814}
  (\bibinfo {year} {2011})}\BibitemShut {NoStop}%
\bibitem [{\citenamefont {Tokman}\ \emph {et~al.}(2015)\citenamefont {Tokman},
  \citenamefont {Erukhimova},\ and\ \citenamefont {Vdovin}}]{Annals}%
  \BibitemOpen
  \bibfield  {author} {\bibinfo {author} {\bibfnamefont {M.~D.}\ \bibnamefont
  {Tokman}}, \bibinfo {author} {\bibfnamefont {M.~A.}\ \bibnamefont
  {Erukhimova}}, \ and\ \bibinfo {author} {\bibfnamefont {V.~V.}\ \bibnamefont
  {Vdovin}},\ }\href@noop {} {\bibfield  {journal} {\bibinfo  {journal} {Annals
  of Physics}\ }\textbf {\bibinfo {volume} {360}},\ \bibinfo {pages} {571}
  (\bibinfo {year} {2015})}\BibitemShut {NoStop}%
\bibitem [{\citenamefont {Tokman}\ \emph {et~al.}(2013)\citenamefont {Tokman},
  \citenamefont {Yao},\ and\ \citenamefont {Belyanin}}]{Tokman_Yao_Belyanin}%
  \BibitemOpen
  \bibfield  {author} {\bibinfo {author} {\bibfnamefont {M.}~\bibnamefont
  {Tokman}}, \bibinfo {author} {\bibfnamefont {X.}~\bibnamefont {Yao}}, \ and\
  \bibinfo {author} {\bibfnamefont {A.}~\bibnamefont {Belyanin}},\ }\href@noop
  {} {\bibfield  {journal} {\bibinfo  {journal} {Phys. Rev. Lett.}\ }\textbf
  {\bibinfo {volume} {110}},\ \bibinfo {pages} {077404} (\bibinfo {year}
  {2013})}\BibitemShut {NoStop}%
\bibitem [{\citenamefont {Blum}(2012)}]{Blum}%
  \BibitemOpen
  \bibfield  {author} {\bibinfo {author} {\bibfnamefont {K.}~\bibnamefont
  {Blum}},\ }\href@noop {} {\emph {\bibinfo {title} {Density Matrix Theory and
  Applications}}}\ (\bibinfo  {publisher} {Springer},\ \bibinfo {address} {New
  York},\ \bibinfo {year} {2012})\BibitemShut {NoStop}%
\bibitem [{\citenamefont {Fain}\ and\ \citenamefont {Khanin}(1969)}]{Fain}%
  \BibitemOpen
  \bibfield  {author} {\bibinfo {author} {\bibfnamefont {V.~M.}\ \bibnamefont
  {Fain}}\ and\ \bibinfo {author} {\bibfnamefont {Y.~I.}\ \bibnamefont
  {Khanin}},\ }\href@noop {} {\emph {\bibinfo {title} {Quantum electronics.
  Basic theory}}}\ (\bibinfo  {publisher} {MIT},\ \bibinfo {address}
  {Cambridge, MA},\ \bibinfo {year} {1969})\BibitemShut {NoStop}%
\bibitem [{\citenamefont {Scully}\ and\ \citenamefont
  {Zubairy}(1997)}]{Scully}%
  \BibitemOpen
  \bibfield  {author} {\bibinfo {author} {\bibfnamefont {M.~O.}\ \bibnamefont
  {Scully}}\ and\ \bibinfo {author} {\bibfnamefont {M.~S.}\ \bibnamefont
  {Zubairy}},\ }\href@noop {} {\emph {\bibinfo {title} {Quantum Optics}}}\
  (\bibinfo  {publisher} {Cambridge University Press},\ \bibinfo {address}
  {Cambridge, New York},\ \bibinfo {year} {1997})\BibitemShut {NoStop}%
\bibitem [{\citenamefont {Kocharovskaya}\ \emph {et~al.}(1992)\citenamefont
  {Kocharovskaya}, \citenamefont {Mandel},\ and\ \citenamefont
  {Radeonychev}}]{Kochar_Mandel_Radion}%
  \BibitemOpen
  \bibfield  {author} {\bibinfo {author} {\bibfnamefont {O.}~\bibnamefont
  {Kocharovskaya}}, \bibinfo {author} {\bibfnamefont {P.}~\bibnamefont
  {Mandel}}, \ and\ \bibinfo {author} {\bibfnamefont {Y.~V.}\ \bibnamefont
  {Radeonychev}},\ }\href@noop {} {\bibfield  {journal} {\bibinfo  {journal}
  {Phys. Rev. A}\ }\textbf {\bibinfo {volume} {45}},\ \bibinfo {pages} {1997}
  (\bibinfo {year} {1992})}\BibitemShut {NoStop}%
\bibitem [{\citenamefont {Brillouin}\ and\ \citenamefont
  {Sommerfeld}(1960)}]{Brillouin}%
  \BibitemOpen
  \bibfield  {author} {\bibinfo {author} {\bibfnamefont {L.}~\bibnamefont
  {Brillouin}}\ and\ \bibinfo {author} {\bibfnamefont {A.}~\bibnamefont
  {Sommerfeld}},\ }\href@noop {} {\emph {\bibinfo {title} {Wave Propagation and
  Group Velocity}}}\ (\bibinfo  {publisher} {Academic Press},\ \bibinfo
  {address} {New York},\ \bibinfo {year} {1960})\BibitemShut {NoStop}%
\bibitem [{\citenamefont {Wang}\ \emph {et~al.}(2000)\citenamefont {Wang},
  \citenamefont {Kuzmich},\ and\ \citenamefont {Dogariu}}]{Wang}%
  \BibitemOpen
  \bibfield  {author} {\bibinfo {author} {\bibfnamefont {L.~J.}\ \bibnamefont
  {Wang}}, \bibinfo {author} {\bibfnamefont {A.}~\bibnamefont {Kuzmich}}, \
  and\ \bibinfo {author} {\bibfnamefont {A.}~\bibnamefont {Dogariu}},\
  }\href@noop {} {\bibfield  {journal} {\bibinfo  {journal} {Nature}\ }\textbf
  {\bibinfo {volume} {406}},\ \bibinfo {pages} {277} (\bibinfo {year}
  {2000})}\BibitemShut {NoStop}%
\bibitem [{\citenamefont {Kryachko}\ \emph {et~al.}(2006)\citenamefont
  {Kryachko}, \citenamefont {Tokman},\ and\ \citenamefont
  {Westerhof}}]{Kryachko}%
  \BibitemOpen
  \bibfield  {author} {\bibinfo {author} {\bibfnamefont {A.~Y.}\ \bibnamefont
  {Kryachko}}, \bibinfo {author} {\bibfnamefont {M.~D.}\ \bibnamefont
  {Tokman}}, \ and\ \bibinfo {author} {\bibfnamefont {E.}~\bibnamefont
  {Westerhof}},\ }\href@noop {} {\bibfield  {journal} {\bibinfo  {journal}
  {Physics of plasmas}\ }\textbf {\bibinfo {volume} {13}},\ \bibinfo {pages}
  {072106} (\bibinfo {year} {2006})}\BibitemShut {NoStop}%
\bibitem [{\citenamefont {Yuen}\ and\ \citenamefont {Chan}(1983)}]{Yuen}%
  \BibitemOpen
  \bibfield  {author} {\bibinfo {author} {\bibfnamefont {H.~P.}\ \bibnamefont
  {Yuen}}\ and\ \bibinfo {author} {\bibfnamefont {V.~W.~S.}\ \bibnamefont
  {Chan}},\ }\href@noop {} {\bibfield  {journal} {\bibinfo  {journal} {Opt.
  Lett.}\ }\textbf {\bibinfo {volume} {8}},\ \bibinfo {pages} {177} (\bibinfo
  {year} {1983})}\BibitemShut {NoStop}%
\bibitem [{\citenamefont {Weiss}(1999)}]{Weiss}%
  \BibitemOpen
  \bibfield  {author} {\bibinfo {author} {\bibfnamefont {U.}~\bibnamefont
  {Weiss}},\ }\href@noop {} {\emph {\bibinfo {title} {Quantum dissipative
  systems}}}\ (\bibinfo  {publisher} {World scientific},\ \bibinfo {address}
  {Singapore},\ \bibinfo {year} {1999})\BibitemShut {NoStop}%
\bibitem [{\citenamefont {Cohen-Tannoudji}\ \emph {et~al.}(1992)\citenamefont
  {Cohen-Tannoudji}, \citenamefont {Dupont-Roc},\ and\ \citenamefont
  {Grynberg}}]{Cohen-Tannoudji}%
  \BibitemOpen
  \bibfield  {author} {\bibinfo {author} {\bibfnamefont {C.}~\bibnamefont
  {Cohen-Tannoudji}}, \bibinfo {author} {\bibfnamefont {J.}~\bibnamefont
  {Dupont-Roc}}, \ and\ \bibinfo {author} {\bibfnamefont {G.}~\bibnamefont
  {Grynberg}},\ }\href@noop {} {\emph {\bibinfo {title} {Quantum Optics}}}\
  (\bibinfo  {publisher} {Atom-Photon Interactions: Basic Processes and
  Applications},\ \bibinfo {address} {Willey, New York},\ \bibinfo {year}
  {1992})\BibitemShut {NoStop}%
\end{thebibliography}%

\appendix

\section{\label{Appendix2} Correlation functions of atomic Langevin operators}
Here we analyse the correlation properties of Langevin operators in atomic equations Eq.~\formref{rho}:
\begin{equation}
\dot{\hat{\rho}}_{mn}
=-\frac{i}{\hslash}\brac{\hat{h}_{mp}\hat{\rho}_{pn}-\hat{\rho}_{mp}\hat{h}_{pn}}+\hat{R}_{mn}+\hat{F}_{mn}.
\label{rhoA}
\end{equation}
The important point is that the properties of noise operators \(\hat{F}_{mn}\) are connected with the properties of relaxation operators \(\hat{R}_{mn}\) \cite{Weiss}.  
  
The model of constant relaxation rates Eq.~\formref{r_mnpq} obtained within the frame of the Markov approximation \cite{Weiss} is equivalent to the \(\delta\)-correlation in time of the noise source:
\begin{equation}
\bracm{\hat{F}_{mn}(\mathbf{r},t)\hat{F}_{pq}(\mathbf{r'},t')}=2D_{mnpq}\brac{\mathbf{r},\mathbf{r'},t}\delta\brac{t-t'}.
\label{dt}
\end{equation} 
This approximation allows one to use so-called generalized Einstein relations \cite{Scully, Cohen-Tannoudji} for calculating the diffusion coefficients \(D_{mnpq}\brac{\mathbf{r},\mathbf{r'},t}\):
\begin{eqnarray}
2D_{mnpg}\brac{\mathbf{r},\mathbf{r'}}=\frac{d}{dt}\bracm{\hat{\rho}_{mn}(\mathbf{r},t)\hat{\rho}_{pq}(\mathbf{r'},t)}-\nonumber\\
-\bracm{\brac{\frac{d}{dt}\hat{\rho}_{mn}(\mathbf{r},t)-\hat{F}_{mn}(\mathbf{r},t)}\hat{\rho}_{pq}(\mathbf{r'},t)}-\nonumber\\
-\bracm{\hat{\rho}_{mn}(\mathbf{r},t)\brac{\frac{d}{dt}\hat{\rho}_{pq}(\mathbf{r'},t)-\hat{F}_{pq}(\mathbf{r'},t)}}.
\label{GER}
\end{eqnarray}
Next, we assume that the action of reservoir on different atoms is independent, so that fluctuations of density matrix operators for different atoms are not correlated \(\bracm{\brac{\hat{\rho}_{mn;j}-\bracm{\hat{\rho}_{mn;j}}}\brac{\hat{\rho}_{pq;i}-\bracm{\hat{\rho}_{pq;i}}}}\propto\delta_{ij}\).
Taking into account also the strict equality, that should be fulfilled for each atom by definition of density matrix operators: \(\hat{\rho}_{mn;j}\hat{\rho}_{pq;j}=\hat{\rho}_{pn;j}\delta_{mq}\), we get for the averaged product of space-dependent density matrix operators the following relation:
\begin{eqnarray}
\bracm{\hat{\rho}_{mn}(\mathbf{r})\hat{\rho}_{pq}(\mathbf{r'})}=\nonumber\\
=\bracm{\hat{\rho}_{mn}(\mathbf{r})}\bracm{\hat{\rho}_{pq}(\mathbf{r'})}+\delta_{mq}\bracm{\hat{\rho}_{pn}(\mathbf{r})}\delta(\mathbf{r}-\mathbf{r'}).
\label{product_rho}
\end{eqnarray}
The expression for the diffusion coefficients Eq.~\formref{GER} with regard to Eqs.~\formref{product_rho},\formref{rho},\formref{r_mnpq} finally takes \(\delta\)-correlated in space form:  
\begin{equation}
D_{mnpq}(\mathbf{r},\mathbf{r'},t)=D_{mnpq}(\mathbf{r},t)\delta(\mathbf{r}-\mathbf{r}'),
\label{dr}
\end{equation}
where 
\begin{eqnarray}
2D_{mnpq}(\mathbf{r},t)=\nonumber\\
=\delta_{mq}\bracm{\hat{R}_{pn}}-\sum_{l}r_{mnql}\bracm{\hat{\rho}_{pl}}-\sum_{k}r_{pqlm}\bracm{\hat{\rho}_{kn}}.
\nonumber
\end{eqnarray}
In a simple case of Eq.~\formref{gamma_mn} the correlation functions of the Langevin sources for the "off-diagonal" operators (\(m\neq n\),\(p\neq q\)) are given by the following expression:
\begin{equation}
2D_{mnpq}(\mathbf{r},t)=\delta_{mq}\brac{\brac{\gamma_{mn}+\gamma_{pq}}\bracm{\hat{\rho}_{pn}}+\bracm{\hat{R}_{pn}}}.
\label{D_mnpq}
\end{equation}
Thus, the autocorrelation function for Langevin operator at some transition \(m-n\) is given by:
\begin{eqnarray}
2D_{mnnm}(\mathbf{r},t)=2\gamma_{mn}\bracm{\hat{\rho}_{nn}}+\bracm{\hat{R}_{nn}}.
\label{D_mnnm}
\end{eqnarray}
The excited coherence at some transition \(\cet{a}-\cet{b}\) corresponds to the non-zero correlations of Langevin sources at the adjacent atomic transitions:
\begin{eqnarray}
2D_{mabm}(\mathbf{r},t)=\brac{\gamma_{am}+\gamma_{bm}-\gamma_{ab}}\bracm{\hat{\rho}_{ba}}.
\label{D_mabm}
\end{eqnarray} 

For the spectral components of the Langevin operators defined as \(\hat{F}_{mn}\brac{\mathbf{r},t}=\int_{-\infty}^{+\infty}{\hat{F}_{mn}(\mathbf{r},\omega)e^{-i\omega t}d\omega}\), 
taking into account Eq.~\formref{dt}, Eq.~\formref{dr}, we get:
\begin{equation}
\bracm{\hat{F}_{mn}(\mathbf{r},\omega)\hat{F}_{pq}(\mathbf{r'},\omega')}=\frac{1}{\pi}D_{mnpq}\brac{\mathbf{r},\omega+\omega'}\delta(\mathbf{r}-\mathbf{r}'),
\end{equation}
where
\begin{equation}
D_{mnpq}\brac{\mathbf{r},\omega}=\frac{1}{2\pi}\int_{-\infty}^{+\infty}{D_{mnpq}(\mathbf{r},t)e^{i\omega t}dt}.
\end{equation}
Under the adiabatic approximation, neglecting slow evolution of populations and amplitude of drive-induced coherence in resonant approximation \(\bracm{\hat{\rho}_{nn}}\approx const\), \(\bracm{\hat{\rho}_{ba}}=\left.{\sigma_{ba}e^{\mp i\omega_d t}}\right|_{b\gtrless a}\), \(\sigma_{ba}\approx const\) the atomic noise operators are \(\delta\)-correlated in frequency and we get from Eq.~\formref{D_mnnm} and Eq.~\formref{D_mabm} the following correlation functions:
\begin{eqnarray}
\bracm{\hat{F}_{mn}(\mathbf{r},\omega)\hat{F}_{nm}(\mathbf{r'},\omega')}=\nonumber\\
\frac{1}{2\pi}\brac{2\gamma_{mn}\bracm{\hat{\rho}_{nn}}+\bracm{\hat{R}_{nn}}}\delta(\omega+\omega')\delta(\mathbf{r}-\mathbf{r}')\nonumber\\
\label{mnnm_omega_A}
\end{eqnarray}
\begin{eqnarray}
\bracm{\hat{F}_{ma}(\mathbf{r},\omega)\hat{F}_{bm}(\mathbf{r'},\omega')}=
\frac{1}{2\pi}\brac{\gamma_{am}+\gamma_{bm}-\gamma_{ab}}\times\nonumber\\
\left.{\sigma_{ba}\delta(\omega+\omega'\mp \omega_d)}\right|_{b\gtrless a}\delta(\mathbf{r}-\mathbf{r}').\nonumber\\
\label{mabm_omega_A}
\end{eqnarray}

\end{document}